\documentclass[12pt]{iopart}
\usepackage{graphicx}
\usepackage{amssymb}

\newcommand{\nn}{\nonumber}
\newcommand{\pa}{\partial}

\begin{document}

\title{Singular statistics revised}
\author{T.~Tudorovskiy, U.~Kuhl, H.-J.~St\"ockmann}
\ead{timur.tudorovskiy@physik.uni-marburg.de}
%\homepage[]{http://www.physik.uni-marburg.de/de/forschung/quantenchaos/ag-startseite.html}
\address{Fachbereich Physik der Philipps-Universit\"at Marburg, Renthof 5,
D-35032, Germany}

\date{\today}

\begin{abstract}
In the paper we analyze the ``singular statistics'' of pseudointegrable \v{S}eba billiards and show that taking into account growing number of resonances one observes the transition from ``semi-Poissonian''-like statistics to Poissonian. This observation is in agreement with an argument that a classical particle does not feel a point perturbation. However, our findings contradict results reported earlier (P.~\v{S}eba, Phys.~Rev.~Lett. \textbf{64}, 1855 (1990)).
\end{abstract}

% 05.45.-a Nonlinear dynamics and chaos
% 05.45.Ac Low-dimensional chaos
% 03.65.Nk Scattering theory
\submitto{New Journal of Physics}

\maketitle

\section{Introduction}

The singular perturbed square billiard, also called \v{S}eba billiard \cite{seb90} by several authors, is one of the key models for quantum chaotic systems. While the unperturbed square billiard with side ratio $1/(\sqrt{5}-1)$ shows Poissonian level-spacings statistics \cite{cas85} and the Sinai billiard, being proved to be a fully chaotic system \cite{sin70}, exhibits GOE-statistics \cite{ber81}, the ``intermediate'' case of a singular perturbed billiard is expected to demonstrate some transient behavior \cite{seb90,seb91,haa91,bog01}. However it was reported, that the billiard with a point perturbation can exhibit ``fully developed quantum chaos'' \cite{seb90,haa91}. It may seem strange, since a point perturbation has almost no influence on the classical phase space of the billiard. Thus it is natural to assume that in the semiclassical limit the billiard with a point perturbation shows a similar statistics to an unperturbed one.

The proposed explanation of the given paradox was based on the argument that for zero-range perturbation for any wavelength one can never reach the limit of the classical billiard with a point-perturbation, since the wavelength is finite while the perturbation radius is zero. Therefore the quantum system becomes chaotic while its classical analog is almost integrable. This argument seemingly was justified experimentally \cite{haa91}.

In the presented report we show that the level-spacings statistics of \v{S}eba billiards actually tends to Poissonian when the number of taken eigenvalues tends to infinity. These findings are in accordance with the intuitive ``classical'' argument given above, but are in contradiction with previous theoretical \cite{seb90,bog01} and experimental \cite{haa91} results. For a narrow window of eigenvalues some conclusions of Refs. \cite{seb90,bog01} remain valid, however one cannot directly apply the results to a wide eigenvalue range. This discrepancy traces back to the procedure applied by the authors to take care of the singularity of the Green function, by replacing the ``bare'' coupling constant by a renormalized one absorbing the infinity. Although this renormalization technique  is standard in quantum electrodynamics, it is not appropriate to compute the spectrum of a \v{S}eba billiard. In this paper we present the suitable renormalization procedure.

\begin{figure}
\begin{center}
\begin{minipage}{40mm}
\begin{tabular}{c}
 \includegraphics[width=35mm]{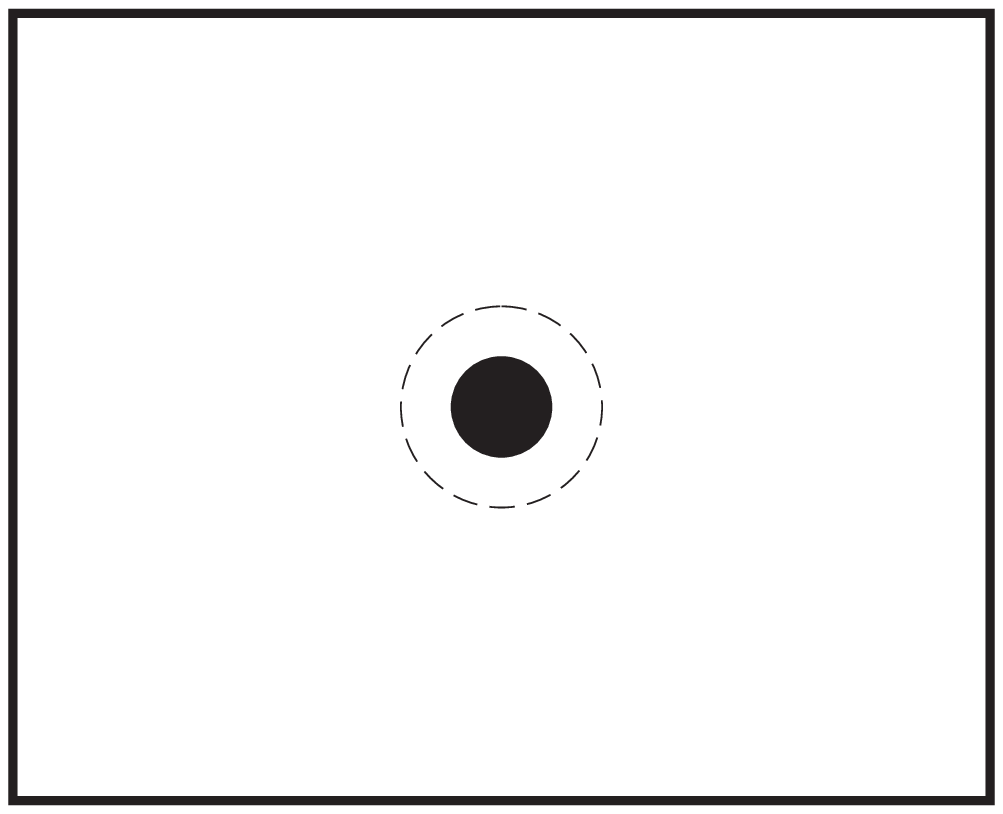} \\
 (a)
\end{tabular}
\end{minipage}
\begin{minipage}{40mm}
\begin{tabular}{c}
 \includegraphics[width=35mm]{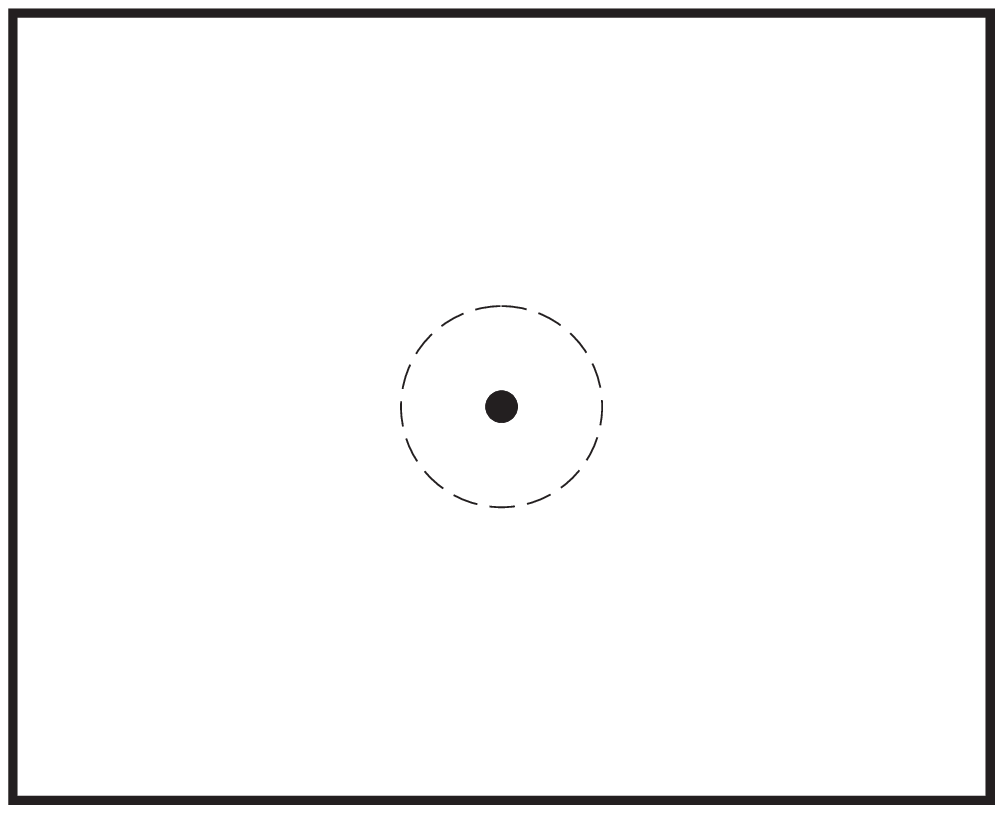} \\
 (b)
\end{tabular}
\end{minipage}
\end{center}
\caption{\label{fig::rb}Physical scatterers of the same scattering length but different radii introduced into the rectangular billiard. Black circle shows the scatterer, the radius of the dashed circle is equal to the scattering length.}
\end{figure}

Let us turn to the experimental microwave \v{S}eba billiard \cite{haa91} and explain why the interpretation of the obtained distributions was erroneous. Theoretically in all cases the scatterer was treated as a point scatterer, which means physically that the characteristic wavelength $\lambda$ of the field inside the cavity is much larger than the radius $a$ of the scatterer. At the same time, the scattering length $\beta$ of the given scatterer is, generally speaking, a free parameter depending on the internal structure of the scatterer (e.\,g. given material of the coating, radius of the metallic core etc). We show below that the influence of the point scatterer is significant when $\lambda\gtrsim\beta$ and vanishes when $\lambda\ll\beta$ in accord with a classical limit. To treat the scatterer as a point perturbation we obligatory should require $a\ll \lambda$. Combining the last two estimations we conclude that to cover experimentally the classical limit of a \v{S}eba billiard one needs to create a scatterer with $a\ll \beta$ (see figure~\ref{fig::rb} (b)).

However, experimentally one often has the situation shown in figure~\ref{fig::rb} (a), i.\,e. $a\sim\beta$. This means that in the regime $\lambda\sim\beta$ the corresponding billiard should be treated as a quantum Sinai billiard, but not a \v{S}eba billiard. In this case the ``classical'' limit with a point perturbation can not be achieved.

\begin{figure}
\begin{center}
\begin{minipage}{40mm}
\begin{tabular}{c}
 \includegraphics[width=40mm]{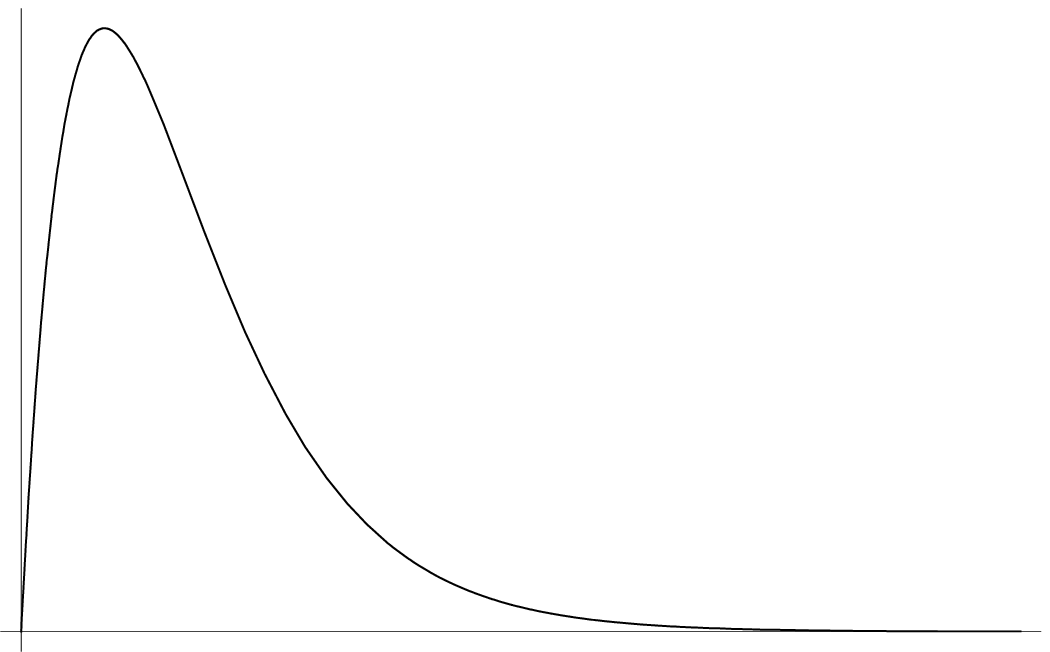} \\
 (a)
\end{tabular}
\end{minipage}
\begin{minipage}{40mm}
\begin{tabular}{c}
 \includegraphics[width=40mm]{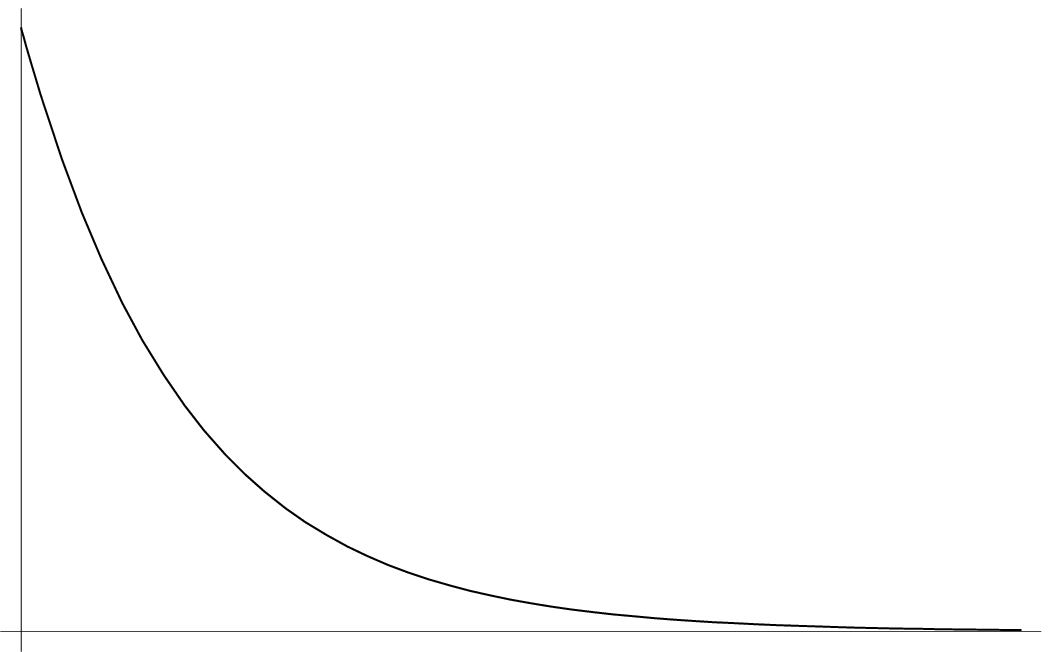} \\
 (b)
\end{tabular}
\end{minipage}
\begin{minipage}{40mm}
\begin{tabular}{c}
 \includegraphics[width=40mm]{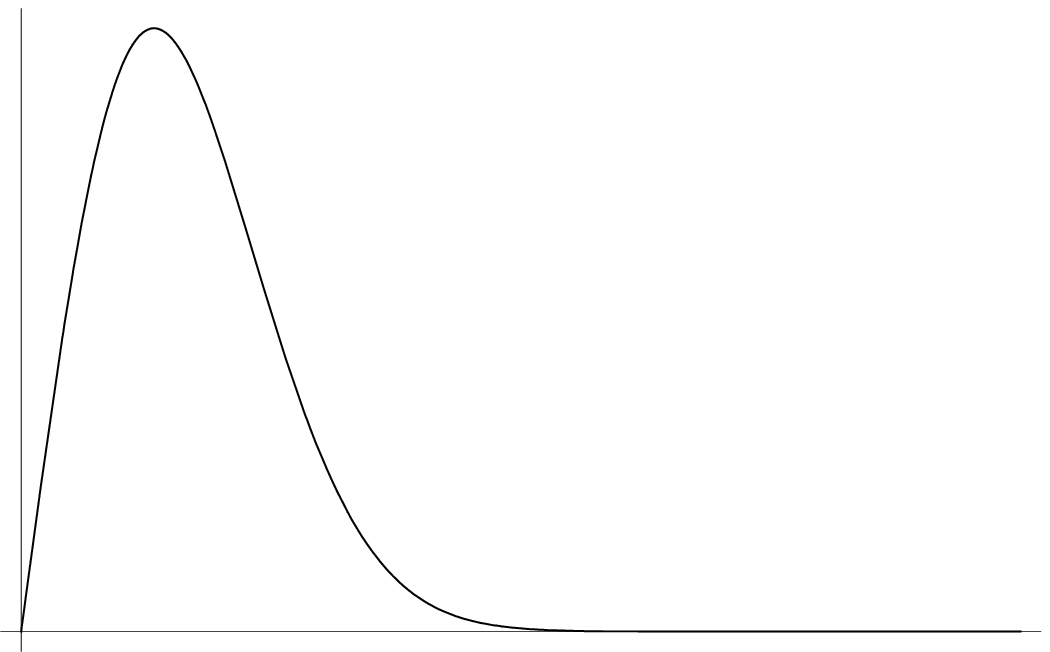} \\
 (c)
\end{tabular}
\end{minipage}
\end{center}
\caption{\label{fig::distrs}Sketch of the level-spacings statistics evolution computed for a given fixed number of eigenvalues of a billiard with a small perturbation whose radius $a$ is significantly smaller than its scattering wavelength $\beta$. The direction from left to right corresponds to increasing eigenvalues. The ``semi-Poissonian'' statistics (a) corresponds to the range $a\ll\beta\lesssim\lambda$, the ``Poissonian'' statistics (b) corresponds to the range $a\ll\lambda\ll\beta$ and the ``GOE'' statistics (c) corresponds to the range $a\sim\lambda\ll\beta$. Here $\lambda$ is the characteristic wavelength. Here $\lambda$ is the characteristic wavelength. The names of distributions are written in parenthesis since the corresponding plotted curves keep the essentials of these statistics but may differ from real distributions.}
\end{figure}

The expected ``experimental'' evolution of the level-spacings statistics computed for a given number of resonances taken at different frequencies is plotted in Fig. \ref{fig::distrs}. It has been assumed that the radius of a small scatterer $a\ll\beta$, but remains finite. Figures 2\,(a), 2\,(b) correspond to the \v{Seba} billiard approximation when the radius of the perturbation can be neglected. Figure 2\,(c) shows the ``GOE'' statistics of the Sinai billiard in the regime where the wavelength of the field is comparable with the radius of the perturbation.

In what follows we take the limit $a\to 0$ which means that figure 2\,(c) can not be reproduced within the framework of the considered approach. Thus we restrict ourself to the \textit{mathematical model of a point perturbation} as it was done by the authors of Refs. \cite{seb90,bog01}.

\section{Point perturbation of the billiard}

Let us now come to the theoretical and numerical study of the \v{S}eba billiard and show why the previous treatments have given incorrect results. Following \cite{dem88,alb88,seb90} let us first introduce a \textit{point perturbation} of the billiard at point $\textbf{R}$. Basically we will construct the ``self-adjoint extension'' of the unperturbed ``Hamiltonian''. This approach was already previously used in \cite{tud08}.

For the unperturbed billiard the eigenfunctions $\psi$ and eigenvalues $k^2=k_n^2$ obey the equation
\begin{equation}
(\Delta+k^2)\psi(\mathbf{r})=0.
\label{eq::unpert}
\end{equation}
If $\psi(\mathbf{R})=0$, the corresponding states do not feel the perturbation, thus these eigenfunctions and the corresponding eigenvalues are identical for the unperturbed and perturbed billiards. Next we assume that the perturbed eigenfunctions $G$ obey the equation
\begin{equation}
(\Delta+k^2)G(\mathbf{r},\mathbf{R};k)=0
\label{eq::pert0}
\end{equation}
outside of the scatterer of the radius $a$. In what follows we assume $a\to 0$. We shall see below that $G$ is nothing but the Green function of the unperturbed system. To recover an appropriate boundary condition at the perturbation point let us consider the asymptotics of the function $G(\mathbf{r},\mathbf{R};k)$ outside of the scatterer when $\mathbf{r}$ tends to $\mathbf{R}$. Rewriting (\ref{eq::pert0}) in cylindrical coordinates we obtain
\begin{equation}
\left[\frac{1}{\rho}\frac{\pa}{\pa\rho}\rho\frac{\pa}{\pa\rho}+\frac{1}{\rho^2}\frac{\pa^2}{\pa\varphi^2}+
k^2\right]G(\mathbf{r},\mathbf{R};k)=0,
\label{eq::cyl1}
\end{equation}
where $\rho=|\mathbf{r}-\mathbf{R}|$ and $\varphi$ is the angle between the vector $\mathbf{r}-\mathbf{R}$ and $x$-axis going along a side of the rectangle.
We require $G(\mathbf{r},\mathbf{R};k)$ to be cylindrically symmetric in the vicinity of the point $\mathbf{r}=\mathbf{R}$, which assumes that the scatterer is cylindrically symmetric. Thus we obtain
\begin{equation}
\left[\frac{1}{\rho}\frac{\pa}{\pa\rho}\rho\frac{\pa}{\pa\rho}+k^2\right]G(\mathbf{r}\to \mathbf{R},\mathbf{R};k)=0.
\label{eq::cyl2}
\end{equation}
The solution of the last equation is
\begin{equation}
G(\mathbf{r}\to \mathbf{R},\mathbf{R};k)=c_1(\mathbf{R};k) J_0(k\rho)+c_2(\mathbf{R};k) Y_0(k\rho),
\label{eq::gas}
\end{equation}
where $J_0$ and $Y_0$ are Bessel functions of the first and the second kind respectively and $c_1, c_2$ are  constants parametrically depending on $\mathbf{R}$ and $k$. The last equality should be understood in asymptotic sense only, since $J_0$ and $Y_0$ do not obey the proper conditions at the outer boundary of the billiard.

Using the asymptotic form of $Y_0(z\to 0)$ \cite{gra80}
\begin{equation}
Y_0(z\to 0)=\frac{2}{\pi}\left[\ln\left(\frac{z}{2}\right)+\gamma\right]+O(z^2\ln\,z),
\label{eq::yas}
\end{equation}
where $\gamma$ is the Euler constant, and the ``identity''
\begin{equation}
\Delta\left(\frac{1}{2\pi}\ln (k|\mathbf{r}-\mathbf{R}|)\right)=\delta(\mathbf{r}-\mathbf{R}),
\label{eq::ln-delta}
\end{equation}
we find that in the limit $a\to 0$ the function $G(\mathbf{r},\mathbf{R};k)$ obeys the equation
\begin{equation}
(\Delta+k^2)G(\mathbf{r},\mathbf{R};k)=\delta(\mathbf{r}-\mathbf{R})
\label{eq::pert}
\end{equation}
if we assume $c_2(\mathbf{R};k)=1/4$. Another choice of the constant $c_2(\mathbf{R};k)$ would only lead to a different normalization.
Taking into account the boundary conditions for the function $G(\mathbf{r},\mathbf{R};k)$ at the outer boundary of the billiard and (\ref{eq::pert}) we conclude that \textit{the perturbed eigenfunctions are the Green functions of the unperturbed billiard}.

We are now going to derive the proper boundary condition at the perturbation point. First we separate the Green function into its regular and singular part, respectively.

\begin{figure}
\begin{center}
 \includegraphics[width=70mm]{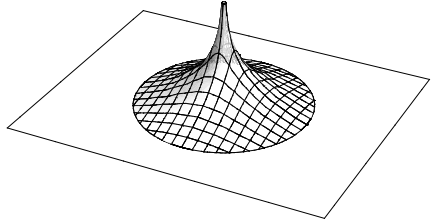}
\end{center}
 \caption{\label{fig::singb}The illustration of the logarithmic singularity of the perturbed eigenfunction}
\end{figure}

Combining (\ref{eq::gas}), (\ref{eq::yas}) we obtain
\begin{equation}
G(\mathbf{r}\to\mathbf{R},\mathbf{R};k)=\frac{1}{2\pi}\ln\left(\frac{\rho}{b}\right)+\xi_b(\mathbf{R};k)+O(z^2\ln\,z),
\label{eq::gexp}
\end{equation}
where $z=k\rho$, $b$ is some arbitrary length, and
\begin{equation}
\xi_b(\mathbf{R};k)=c_1(\mathbf{R};k)+\frac{1}{2\pi}\left[\ln\left(\frac{k b}{2}\right)+\gamma\right].
\label{eq::xiexp}
\end{equation}
Here $\xi_b(\mathbf{R};k)$ is the \textit{renormalized} Green function.
In figure \ref{fig::singb} we illustrate the singularity of the Green function near the perturbation point.

Let us now consider the rectangle with a small pricked circle of radius $a$ whose center is situated at the point $\mathbf{R}$. We denote it $\Omega_a$. Then we consider the linear space of functions consisting of two subspaces: \textit{(1)} the subspace of functions $f^{(1)}(\mathbf{r},\mathbf{R})$ vanishing at the outer boundary and possessing the asymptotics
\begin{equation}
f^{(1)}(\mathbf{r}\to \mathbf{R})=B\left[\frac{1}{2\pi}\ln\left(\frac{\rho}{b}\right)+\xi\right]+O(z^2\ln\,z),
\label{eq::gcalas}
\end{equation}
where $B$ and $\xi$ are some constants, and \textit{(2)} the subspace of regular functions $f^{(2)}(\mathbf{r})$ vanishing at the outer boundary of the billiard such that $f^{(2)}(\mathbf{R})=0$. $G(\textbf{r},\textbf{R};k)$ belongs to the subspace \textit{(1)} because of its asymptotic behavior (\ref{eq::gexp}). The unperturbed eigenfunctions $\psi(\textbf{r})$ belong to the subspace \textit{(2)}. Now we study the action of the the operator $-\Delta=-\nabla^2$  on the space of functions defined above. The requirement of hermicity gives
\begin{eqnarray}
\lim_{a\to 0}\left[
\langle f^{(i)}|-\Delta|f^{(j)}\rangle_a-
\langle f^{(j)}|-\Delta|f^{(i)}\rangle^*_a
\right]=0,\label{eq::herm1}
\end{eqnarray}
where $f^{(i)}$, $f^{(j)}$ with $i,\,j=1,2$ are arbitrary functions taken from the  subspaces \textit{(1)} and \textit{(2)}, respectively, and
\begin{equation}
\langle f^{(i)}|-\Delta|f^{(j)}\rangle_a=\int_{\Omega_a}d^2r {f^{(i)}}^*(-\Delta f^{(j)}).
\end{equation}
By means of the Green's theorem we obtain
\begin{eqnarray}\label{eq:greentheorem}
\fl
\langle f^{(i)}|-\Delta| f^{(j)}\rangle_a-\langle f^{(j)}|-\Delta| f^{(i)}\rangle^*_a=
\int_{\Omega_a}d^2r \nabla\left( f^{(j)}\nabla {f^{(i)}}^*- {f^{(i)}}^*\nabla f^{(j)}\right)=\nonumber\\
\fl
=2\pi a \left[ {f^{(i)}}^*(\pa f^{(j)}/\pa\rho)-f^{(j)}(\pa f^{(i)}/\pa\rho)^*\right]_{\rho=a}.
\label{eq::qfdiff}
\end{eqnarray}
Equation (\ref{eq:greentheorem}) shows that (\ref{eq::herm1}) holds automatically if $f^{(i)}$ and $f^{(j)}$ both belong to the subspace \textit{(2)}. Assume now that $i=1$ and $j=2$ or vice versa. Using the asymptotics (\ref{eq::gexp}) we see that (\ref{eq::herm1}) again holds automatically for any $f^{(1)}$ and $f^{(2)}$. Thus the case $i,j=1$ implies the only nontrivial condition superimposed by the hermicity requirement of the constructed operator. Let us take two functions $f^{(1)}_1$ and $f^{(1)}_2$ with asymptotic expansions
\begin{equation}
f^{(1)}_i(\mathbf{r},\mathbf{R})=B_i\left[\frac{1}{2\pi}\ln\left(\frac{\rho}{b}\right)+\xi_i\right]+O(z^2\ln\,z).
\label{eq::gcalas1}
\end{equation}
Substituting (\ref{eq::gcalas1}) into (\ref{eq::qfdiff}) we find
\begin{eqnarray}
\lim_{a\to 0}\left[\langle f^{(1)}_1|-\Delta|f^{(1)}_2\rangle_a-\langle f^{(1)}_2|-\Delta| f^{(1)}_1\rangle^*_a\right]=
B_1^*B_2(\xi_1^*-\xi_2)=0.
\label{eq::xi1-xi2}
\end{eqnarray}
The equality (\ref{eq::xi1-xi2}) must hold for any values of $B_i,\,\xi_i$. This leads to the conclusion that for all functions from the subspace \textit{(2)} the constant $\xi$ in (\ref{eq::gcalas}) is \textit{real} and \textit{the same}.
Let us chose a certain value $\xi=-D$. Then the boundary condition at the perturbation point reads:
\begin{equation}
\xi+D=0.
\label{eq::spaceD}
\end{equation}
provided that the length $b$ is fixed. Comparing (\ref{eq::gexp}) with (\ref{eq::gcalas}) we find that for the perturbed eigenfunctions $G(\mathbf{r},\mathbf{R};k)$ the constant $\xi$ is equal to $\xi_b(\mathbf{R};k)$. Then (\ref{eq::spaceD}) gives
\begin{equation}
\xi_b(\mathbf{R};k)+D=0.
\label{eq::xisp}
\end{equation}

Substituting (\ref{eq::xiexp}) into (\ref{eq::xisp}) we obtain
\begin{equation}
c_1(\mathbf{R};k)+\frac{1}{2\pi}\left[\ln\left(\frac{kb}{2}\right)+\gamma\right]+D=0.
\label{eq::xisp1}
\end{equation}
The proper boundary condition cannot depend on the arbitrary length $b$, but rather should depend on a parameter, characterizing the inner structure of the perturbation. Therefore the length $b$ should be canceled in (\ref{eq::xisp1}) by a proper choice of $D$. This can be achieved by the following choice:
\begin{equation}
D=\frac{1}{2\pi}\ln\left(\frac{\beta}{b}\right),
\label{eq::D}
\end{equation}
where $\beta$ is the \textit{scattering length of the perturbation}. The value of the length $\beta$ can not be obtained from the consideration above since we cut out the area containing the perturbation, thereby loosing the information on it. Thus we draw the conclusion that the scattering length is \textit{the only parameter describing the perturbation} in the limit $ka\ll 1$. Substituting (\ref{eq::D}) in (\ref{eq::xisp}) and using (\ref{eq::xiexp}), we find
\begin{equation}\label{eq:sec}
\xi_\beta(\mathbf{R};k)=c_1(\mathbf{R};k)+\frac{1}{2\pi}\left[\ln\left(\frac{k\beta}{2}\right)+\gamma\right]=0.
\label{eq::xisp2}
\end{equation}
The perturbed part of the spectrum may now be obtained from the solutions $k^2=k^2_n$ of (\ref{eq::xisp2}).

Replacing $b$ by $\beta$, $k$ by $k_n$ in (\ref{eq::gexp}) and using the equality $\xi_\beta(\mathbf{R},k_n)=0$ we find the asymptotic expansion of the perturbed eigenfunction corresponding to the eigenvalue $k_n^2$:
\begin{eqnarray}
G(\mathbf{r}\to\mathbf{R},\mathbf{R};k_n)=\frac{1}{2\pi}\ln\left(\frac{\rho}{\beta}\right)+O(z^2\ln\,z).
\label{eq::gexp2}
\end{eqnarray}
The leading term of the asymptotics (\ref{eq::gexp2}) becomes zero when $\rho=\beta$. This fact can be used to determine experimentally a scattering length of a given perturbation.

We note that the proper definition of the scattering length was missing in Refs. \cite{seb90,exn97}. This has led to the deficiencies discussed above. However in the monograph \cite{dem88}, devoted to point perturbations, the scattering length in two-dimensional problems was properly introduced.

Several conclusions on the level-spacings distribution can be drawn already from (\ref{eq::xisp2}). Indeed, from (\ref{eq::gexp}) and (\ref{eq::xiexp}) we conclude that $c_1(\mathbf{R};k)$ has the same poles as the Green function of the unperturbed billiard. From (\ref{eq::xisp2}) we obtain that when $k$ tends to infinity, the eigenvalues of the perturbed billiard approach the eigenvalues of the unperturbed one. Indeed, close to the eigenvalue $k_n$ the function $c_1(\textbf{R},k)$ may be approximated by $\textrm{const}/(k^2-k_n^2)$ whence follows:
\begin{equation}
\frac{\textrm{const}}{k^2-k_n^2}+\eta_\beta(k)=0,\qquad
\eta_\beta(k)=\frac{1}{2\pi}\ln\left(\frac{k\beta}{2}\right)\gg 1.
\end{equation}
Then $k^2-k_n^2=-2\pi\,\textrm{const}/\ln (k\beta/2)$. In the limit $k\to\infty$ we recover the original spectrum of the billiard! Since the statistics of inter-level spacings for the unperturbed rectangular billiard with chosen side ratio is Poissonian \cite{cas85}, we conclude that \textit{the same statistics for high-lying eigenvalues of the \v{S}eba billiard is also Poissonian}. This is the most important conclusion of the paper, which contradicts the prediction given in \cite{seb90}.

\section{Ewald's representation of the renormalized Green function}

For an explicit  calculation of the spectrum of the perturbed billiard from (\ref{eq:sec}) an expression for $\xi_\beta(\textbf{R};k)$ is needed, which for the general cases by no means is a trivial task. For the rectangle it can be obtained by an application of Ewald's method. The derivation is technical, and anybody not interested in the details may proceed directly to (\ref{eq::xiewald}).

In the paper \cite{bog01} in (\ref{eq::xisp2}) the logarithmic dependence of the dimensionless scattering strength $\eta_\beta(k)$ was missed. Therefore the proper spectral statistics of the \v{S}eba billiard still has to be computed. In this section we explain the numerical procedure, based on Ewald's method \cite{ewa21,lin98,lin99,duf01,pap99,mor06}, which allows us to compute the renormalized Green function and then, from (\ref{eq::xisp2}), the spectral statistics.

We start from the eigenfunction representation of the Green function for the free billiard
\begin{eqnarray}
G(\mathbf{r},\mathbf{R};k)=\sum_{n=1}^{\infty}\sum_{m=1}^{\infty}
\frac{\psi_{nm}(x,y)\psi_{nm}(x',y')}{k^2-E_{nm}},
\label{eq::gser}
\end{eqnarray}
where $\mathbf{R}=(x',y')$,
\begin{eqnarray}
\psi_{nm}(x,y)=\frac{2}{\sqrt{d_x d_y}}\sin\left(\frac{\pi n x}{d_x}\right)\sin\left(\frac{\pi m y}{d_y}\right),\\
E_{nm}=\left(\frac{\pi n}{d_x}\right)^2+\left(\frac{\pi m}{d_y}\right)^2,
\end{eqnarray}
$d_x$ and $d_y$ are the two sides of the rectangle. When $x\to x'$ and $y\to y'$ then the series (\ref{eq::gser}) diverges logarithmically. This is just another manifestation of the well-known singularity of the  Green function for $\mathbf{r}\to \mathbf{R}$, see (\ref{eq::gexp}), which is a local feature and does not depend on outer boundary conditions. This suggests that the eigenfunctions representation is not the appropriate choice to compute the renormalized Green function, but that the images representation \cite{duf01} might be preferable.

\begin{figure}
\begin{center}
 \includegraphics[width=70mm]{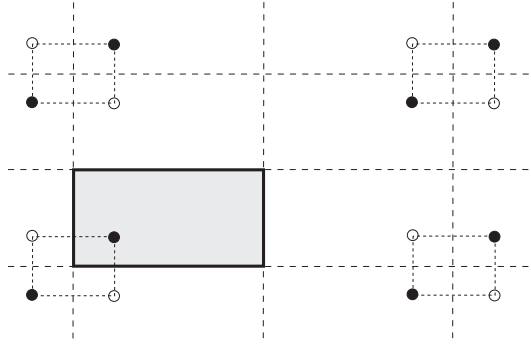}
\end{center}
 \caption{\label{fig::ewald}The illustration of the sources distribution used in the image representation of the Green function}
\end{figure}

The images representation of the Green function reads
\begin{eqnarray}
G(\mathbf{r},\mathbf{R};k)=
\sum_{n,m=-\infty}^\infty \sum_{s_1,s_2=0}^1
(-1)^{s_1+s_2}G_f(\mathbf{r},\mathbf{R}_{s_1 s_2}+\mathbf{R}_{n m};k),
\label{eq::image}
\end{eqnarray}
where
\begin{eqnarray}
\mathbf{R}_{s_1 s_2}=\Bigl((-1)^{s_1}x',(-1)^{s_2}y'\Bigr),\quad
\mathbf{R}_{n m}=(2 n d_x, 2 m d_y),
\end{eqnarray}
and $G_f(\mathbf{r},\mathbf{R};k)$ is the Green function for the two-dimensional plane (see figure \ref{fig::ewald}). The images representation is much better suited to compute the renormalized Green function, since when $\mathbf{r}$ tends to $\mathbf{R}$, the term $n=m=0,\,s_1=s_2=0$ is the only one in the representation logarithmically tending to infinity. Now the divergency can be subtracted analytically. However the image representation does not solve the problem yet since it converges absolutely only if $\textrm{Im}\,k>0$. To overcome this obstacle Ewald \cite{ewa21} proposed the dual representation keeping features of the images as well as the eigenmodes representation.

\begin{figure}
\begin{center}
\begin{minipage}{60mm}
\begin{tabular}{c}
 \includegraphics[width=60mm]{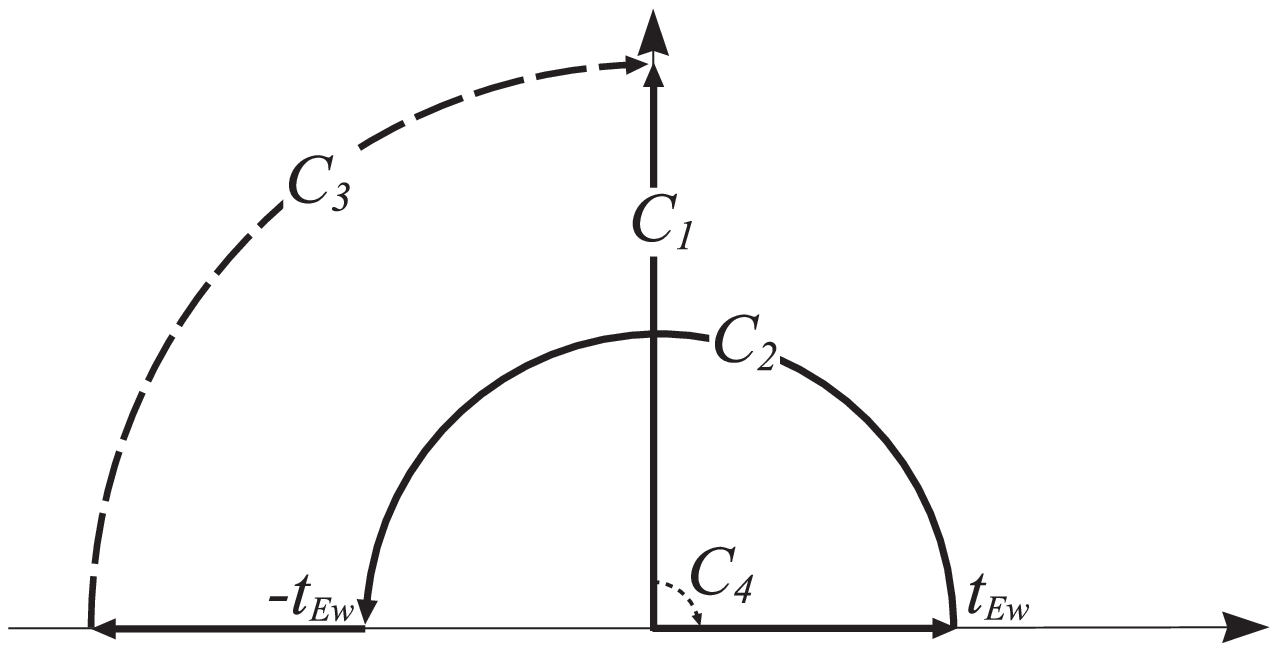}\\
 (a)
\end{tabular}
\end{minipage}
\begin{minipage}{40mm}
\begin{tabular}{c}
 \includegraphics[height=30mm]{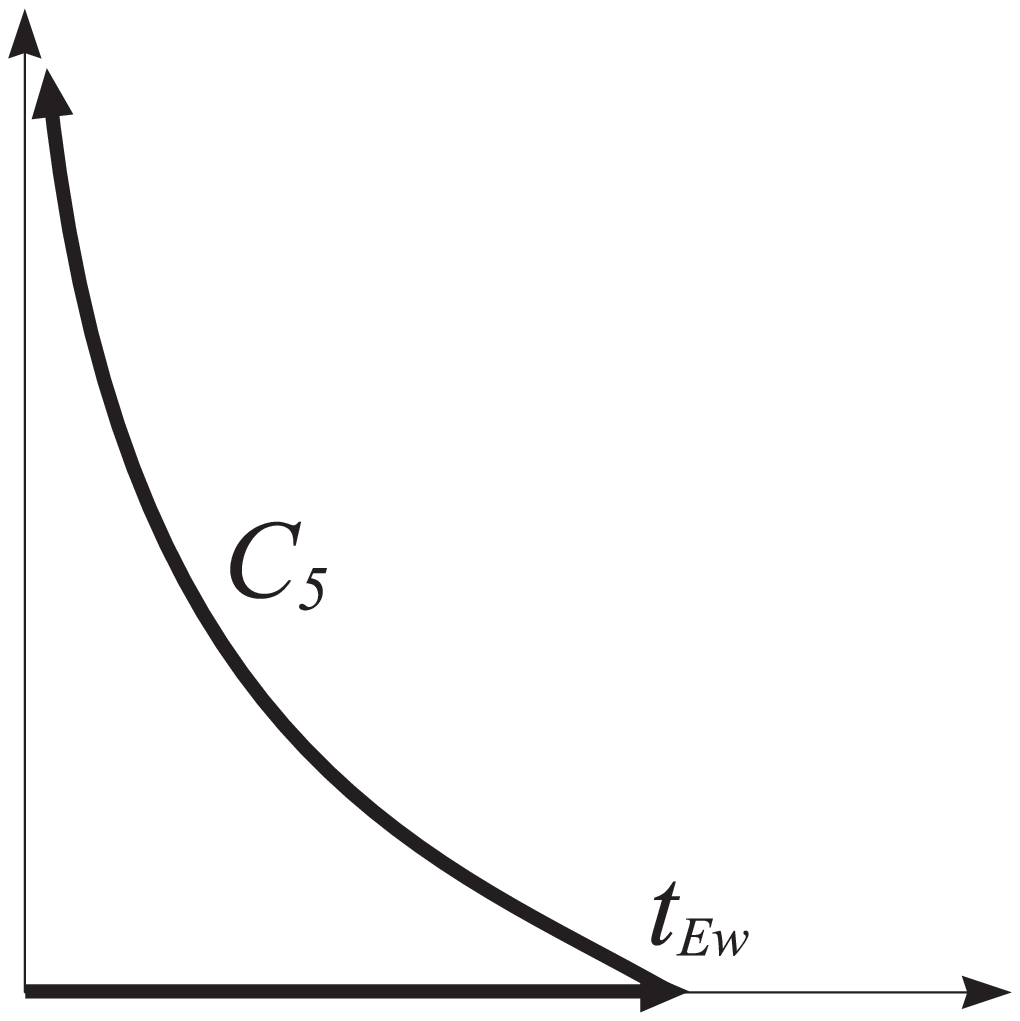}\\
 (b)
\end{tabular}
\end{minipage}
\end{center}
\caption{\label{fig::contour}The contours of the integration for the free Green function $G_f$ (a) and
the contour of the integration for the Ewald's representation of $G_f$ (b)}
\end{figure}

Below we follow the works \cite{lin98,lin99,duf01}. Let us first find a convenient representation for $G_f$. To this end we consider the following initial-value problem for the function $g(t;\mathbf{r},\mathbf{R},k)$
\begin{eqnarray}
\frac{\pa g}{\pa t}=(\Delta+k^2)g, \quad g(t=0;\mathbf{r},\mathbf{R},k)=\delta(\mathbf{r}-\mathbf{R}).
\label{eq::ginit}
\end{eqnarray}
Then $G_f$ can be written as
\begin{equation}
G_f(\mathbf{r},\mathbf{R};k)=-\int_C g(t;\mathbf{r},\mathbf{R},k).
\end{equation}
The contour $C$ should start at $t=0$ and tend to infinity in such a way that $g$ tends to zero. Obviously
\begin{equation}
g(t;\mathbf{r},\mathbf{R},k)=e^{k^2 t}K(t;\mathbf{r},\mathbf{R}),
\end{equation}
where the \textit{heat kernel} $K(t;\mathbf{r},\mathbf{R})$ can be found by a separation of variables
\begin{equation}
K(t;\mathbf{r},\mathbf{R})=\frac{1}{4\pi t}e^{-(\mathbf{r}-\mathbf{R})^2/(4t)}.
\end{equation}
Finally we obtain
\begin{equation}
G_f(\mathbf{r},\mathbf{R};k)=-\int_C \frac{dt}{4\pi t} \exp\left(k^2 t-\frac{(\mathbf{r}-\mathbf{R})^2}{4t}\right).
\label{eq::gint}
\end{equation}
The simplest contour of the integration is the imaginary half-line going from zero to $i\infty$ (contour $C_1$ in figure~\ref{fig::contour} (a)\,). Using this contour we conclude (\cite{gra80}, Entry B.187(2)) that
\begin{equation}\label{eq:reentry}
G_f(\mathbf{r},\mathbf{R};k)
=-\frac{i}{4}H_0^{(1)}(k|\mathbf{r}-\mathbf{R}|),
\label{eq::hank}
\end{equation}
where $H_0^{(1)}$ is the Hankel function of the first kind. However, the asymptotic behavior (\ref{eq::gexp}) becomes hidden in this representation. To recover the asymptotic behavior we will use another contour of the integration. We see that for small values of $|t|$ the best convergency provides an interval lying on the real axis from zero to some positive value $t_{Ew}$ (see figure~\ref{fig::contour} (a)\,). We will call this value the \textit{Ewald parameter}.  From the other side the best convergency for large values of $|t|$ would provide the half-line going from some negative value (we choose it to be equal to $-t_{Ew}$) to $-\infty$. What remains is to connect these parts to make a contour. We connect them by a half-circle $C_2$. Finally the constructed contour is equivalent to $C_1$ since the integrals along the quarter-circles $C_3$, $C_4$ (plotted by dashed lines in figure~\ref{fig::contour} (a)\,) tend to zero when the radius of $C_3$ tends to infinity and the radius of $C_4$ tends to zero.

Using the constructed contour we can obtain the asymptotics of the type (\ref{eq::gexp}) from the representation (\ref{eq::gint}). The developed technic will be used further to compute the renormalized Green function. We write
\begin{eqnarray}
G_f(\mathbf{r},\mathbf{R};k)=G_f^{(1)}(\mathbf{r},\mathbf{R};k)+G_f^{(2)}(\mathbf{r},\mathbf{R};k),
\end{eqnarray}
where
\begin{eqnarray}
G_f^{(1)}(\mathbf{r},\mathbf{R};k)=-\int_0^{t_{Ew}} g(t;\mathbf{r},\mathbf{R},k)dt,\nn \\
G_f^{(2)}(\mathbf{r},\mathbf{R};k)=
-\int_{C_2} g(t;\mathbf{r},\mathbf{R},k)dt
-\int_{-t_{Ew}}^{-\infty} g(t;\mathbf{r},\mathbf{R},k)dt.
\end{eqnarray}
Introducing the notations $u=(\mathbf{r}-\mathbf{R})^2/(4t_{Ew}),\,v=k^2 t_{Ew}$, we write $G_f^{(1)}$ as follows
\begin{equation}\label{eq:Gf1}
G_f^{(1)}(\mathbf{r},\mathbf{R};k)=-\int_0^1 \frac{dt}{4\pi t}e^{v t-u/t}.
\label{eq::g1f-uv}
\end{equation}
When $\mathbf{r}$ tends to $\mathbf{R}$, then $u$ tends to zero. To compute the asymptotics of (\ref{eq::g1f-uv}) for $u\to 0$ we make the following transformations:
\begin{eqnarray}
G_f^{(1)}(\mathbf{r}\to\mathbf{R},\mathbf{R};k)=
-\int_0^1 \frac{dt}{4\pi t} \left(e^{v t}-1+1\right)e^{-u/t}\nn\\
\qquad\simeq\int_0^1 \frac{dt}{4\pi t} \left(1-e^{v t}\right)
-\int_u^\infty\frac{dt}{4\pi t} e^{-t}\nn\\
\qquad=\int_0^1 \frac{dt}{4\pi t} \left(1-e^{v t}\right)-\int_u^1\frac{dt}{4\pi t} e^{-t}-\int_1^\infty\frac{dt}{4\pi t} e^{-t}\nn\\
\qquad
\simeq\int_0^1\frac{dt}{4\pi t} (1-e^{-t})-\int_1^\infty\frac{dt}{4\pi t}e^{-t}
+\int_0^1 \frac{dt}{4\pi t} \left(1-e^{v t}\right)+\frac{1}{4\pi}\ln\,u\nn\\
\qquad
=\frac{1}{4\pi}[g_0(v)+\gamma+\ln\,u],
\label{eq::u0}
\end{eqnarray}
where
\begin{eqnarray}
g_0(v)=\int_0^1 \frac{dt}{t} \left(1-e^{v t}\right),\quad
\gamma=\int_0^1\frac{dt}{t} (1-e^{-t})-\int_1^\infty\frac{dt}{t}e^{-t}
\end{eqnarray}
(see \cite{gra80} (Entry 8.367.12)). Since $g_0(v)$ has no singularity at $t=0$ it can be written as
\begin{eqnarray}
g_0(v)=\int_0^{1} \frac{dt}{t} \left(1-e^{-v t}\right)-\int_{C_0}\frac{dt}{t} \left(1-e^{v t}\right),
\label{eq::g0}
\end{eqnarray}
where $C_0$ is a half-circle of the unit radius.

Since $G_f^{(2)}$ has no singularity when $\mathbf{r}\to\mathbf{R}$, the leading term of its asymptotics is
\begin{eqnarray}
G_f^{(2)}(\mathbf{R},\mathbf{R};k)=-\int_{C_0}\frac{dt}{4\pi t}e^{v t}-\int_1^\infty \frac{dt}{4\pi t}e^{-vt}.
\label{eq::Gf2}
\end{eqnarray}
Using (\ref{eq::u0}), (\ref{eq::g0}), (\ref{eq::Gf2}) we find
\begin{eqnarray}
G_f^{(1)}(\mathbf{r}\to\mathbf{R},\mathbf{R};k)+G_f^{(2)}(\mathbf{R},\mathbf{R};k)\nn\\
\qquad\simeq\frac{1}{4\pi}\left(\ln\,u+
\gamma+\int_0^{1} \frac{dt}{t} \left(1-e^{-v t}\right)
-\int_1^\infty \frac{dt}{t}e^{-vt}\right)-\frac{i}{4}.
\label{eq::g0g2}
\end{eqnarray}
The last equality does not depend on the choice of $t_{Ew}$. To prove it we can take $v$ as the independent parameter, then $u=k^2(\mathbf{r}-\mathbf{R})^2/(4v)$. Differentiation of the last equality with respect to $v$ gives zero. The reasonable choice of $v$ should not lead to exponentially large values of $g_0(v)$. Indeed due to (\ref{eq::g0g2}) such a large contribution is somehow artificial since it is annihilated by $G_f^{(2)}$. Thus it is natural to take $v=1$. Then (\ref{eq::g0g2}) gives
\begin{eqnarray}
G_f(\mathbf{r}\to\mathbf{R},\mathbf{R};k)\simeq \frac{1}{2\pi}\left[\ln\left(\frac{k|\mathbf{r}-\mathbf{R}|}{2}\right)+\gamma\right]-\frac{i}{4}.
\label{eq::Gf}
\end{eqnarray}
This calculation has demonstrated that it is important to divide the free Green function in two parts: $G^{(1)}_f$ and $G^{(2)}_f$. The first part describes the space singularity, and the second part makes the contribution into the regular part of the asymptotics.

Let us turn now to the image representation (\ref{eq::image}) of the Green function of the rectangular cavity.
We again write the free Green function $G_f$ in the form
\begin{equation}
G_f(\mathbf{r},\mathbf{R};k)=G_f^{(1)}(\mathbf{r},\mathbf{R};k)+G_f^{(2)}(\mathbf{r},\mathbf{R};k),
\label{eq::g1g2}
\end{equation}
where $G_f^{(1)}$ is defined by (\ref{eq::g1f-uv}) and
\begin{equation}
G_f^{(2)}(\mathbf{r},\mathbf{R};k)=-\int_{C_5} \frac{dt}{4\pi t} \exp\left(k^2 t-\frac{(\mathbf{r}-\mathbf{R})^2}{4t}\right).
\end{equation}
The best choice of the contour $C_5$ will be discussed later. We shall see that the choice used to compute the asymptotics (\ref{eq::Gf}) does not fit. Substituting (\ref{eq::g1g2}) into (\ref{eq::image}) we obtain
\begin{equation}
G(\mathbf{r},\mathbf{R};k)=G^{(1)}(\mathbf{r},\mathbf{R};k)+G^{(2)}(\mathbf{r},\mathbf{R};k),
\end{equation}
where
\begin{eqnarray}
G^{(1)}(\mathbf{r},\mathbf{R};k)
=\sum_{n,m=-\infty}^\infty \sum_{s_1,s_2=0}^1
(-1)^{s_1+s_2}G_f^{(1)}(\mathbf{r},\mathbf{R}_{s_1 s_2}+\mathbf{R}_{n m};k),\label{eq::g1}\\
G^{(2)}(\mathbf{r},\mathbf{R};k)
=\sum_{n,m=-\infty}^\infty \sum_{s_1,s_2=0}^1
(-1)^{s_1+s_2}G_f^{(2)}(\mathbf{r},\mathbf{R}_{s_1 s_2}+\mathbf{R}_{n m};k).
\label{eq::g2}
\end{eqnarray}
To improve the convergency of series for $G^{(2)}$ we use the identity
\begin{eqnarray}
\sum_{n=-\infty}^\infty e^{-(x-2nd_x)^2/(4t)}=
\frac{\sqrt{\pi t}}{d_x}\sum_{n=-\infty}^\infty e^{i\pi n x/d_x-\pi^2 n^2 t/d_x^2},
\end{eqnarray}
which can be proved by applying the Poisson sum rule \cite{duf01} to the function
\begin{equation}
f(x)=e^{-x^2/(4t)}.
\end{equation}
Performing the resummation we obtain
\begin{eqnarray}
\fl
G^{(2)}(\mathbf{r},\mathbf{R};k)\nn\\
\fl
\qquad=-\frac{1}{4d_x d_y}\sum_{n,m=-\infty}^\infty e^{i\pi n x/d_x+i\pi m y/d_y}
\sum_{s_1,s_2=0}^1(-1)^{s_1+s_2}
\int_{C_5}e^{[k^2-(\pi n/d_x)^2-(\pi m/d_y)^2]t}dt.
\end{eqnarray}
Now we see that the integral over $C_5$ should converge for positive as well as for negative values of the real part of $k^2-(\pi n/d_x)^2-(\pi m/d_y)^2$ provided that $\textrm{Im}\,k>0$. Therefore we have to assume $t\to i\infty$ along the contour $C_5$. This leads to the choice of the contour shown in figure~\ref{fig::contour} (b).

Performing the integration we obtain
\begin{eqnarray}
G^{(2)}(\mathbf{r},\mathbf{R};k)=\frac{1}{4d_x d_y}
\sum_{n,m=-\infty}^\infty \frac{e^{[k^2-(\pi n/d_x)^2-(\pi m/d_y)^2]t_{Ew}}}
{k^2-(\pi n/d_x)^2-(\pi m/d_y)^2}e^{i\pi n x+i\pi m y}\nn\\
\times\sum_{s_1,s_2=0}^1 (-1)^{s_1+s_2}
e^{-i\pi n(-1)^{s_1}x'/d_x-i\pi m(-1)^{s_2}y'/d_y}
.
\end{eqnarray}
Summarizing over $s_1,\,s_2$ we get
\begin{eqnarray}
\fl
\sum_{s_1,s_2=0}^1 (-1)^{s_1+s_2}e^{-i\pi n(-1)^{s_1}x'/d_x-i\pi m(-1)^{s_2}y'/d_y}
=-4\sin\left(\frac{\pi n x'}{d_x}\right)\sin\left(\frac{\pi m y'}{d_y}\right).
\end{eqnarray}
Now we can perform summations over $n$ and $m$:
\begin{eqnarray}\label{eq:G2}
\fl
G^{(2)}(\mathbf{r},\mathbf{R};k)=-\frac{1}{d_x d_y}\nn\\
\fl\qquad
\times
\sum_{n,m=-\infty}^\infty \sin\left(\frac{\pi n x'}{d_x}\right)\sin\left(\frac{\pi m y'}{d_y}\right)
\frac{e^{[k^2-(\pi n/d_x)^2-(\pi m/d_y)^2]t_{Ew}}}{k^2-(\pi n/d_x)^2-(\pi m/d_y)^2}
e^{i\pi n x/d_x+i\pi m y/d_y}=\nn\\
\fl\qquad
=\frac{4}{d_x d_y}
\sum_{n,m=1}^\infty
\sin\left(\frac{\pi n x}{d_x}\right)\sin\left(\frac{\pi m y}{d_y}\right)
\sin\left(\frac{\pi n x'}{d_x}\right)\sin\left(\frac{\pi m y'}{d_y}\right)\nn\\ \times
\frac{e^{[k^2-(\pi n/d_x)^2-(\pi m/d_y)^2]t_{Ew}}}{k^2-(\pi n/d_x)^2-(\pi m/d_y)^2}.
\label{eq::g2em}
\end{eqnarray}
Formulas (\ref{eq::g1f-uv}), (\ref{eq::g1}) and (\ref{eq::g2em}) give the Ewald representation of the Green function for the rectangular billiard. The integral in (\ref{eq::g1f-uv}) has to be computed numerically. Now we can recapitulate the advantages of the Ewald's method. First of all, both series $G^{(1)}$ and $G^{(2)}$ are exponentially convergent. Thus we can take the analytic continuation and choose real $k$. Second, we have separated the part $G^{(1)}$ responsible for the space singularity from the part $G^{(2)}$ responsible for the poles information. Indeed, $G^{(2)}$ exponentially converges even when $\mathbf{r}=\mathbf{R}$. Third, only in $G^{(1)}$ there is the term corresponding to $s_1=s_2=0, n=m=0$ which asymptotically tends to infinity when $\mathbf{r}\to\mathbf{R}$. The rest of the series is exponentially convergent. The last observation allows to compute the renormalized Green function.

Though the Poisson resummation is a common tool used to get the Ewald's representation of the Green function \cite{ewa21,lin98,lin99,duf01,mor06}, one can avoid it and obtain formulas (\ref{eq::g1}) and (\ref{eq::g2em}) easier (see the Appendix for details).

Using the asymptotic expansion (\ref{eq::u0}) we obtain the \textit{exact Ewald representation for the renormalized Green function}:
\begin{eqnarray}
\fl
\xi_\beta(\mathbf{R};k)=\frac{1}{4\pi}[g_0(k^2 t_{Ew})+\gamma]+\nn\\
\fl\qquad
+\sum_{n,m=-\infty}^\infty \sum_{s_1,s_2=0}^1 (1-\delta_{n,0}\delta_{m,0}\delta_{s_1,0}\delta_{s_2,0})
(-1)^{s_1+s_2}G_f^{(1)}(\mathbf{R},\mathbf{R}_{s_1 s_2}+\mathbf{R}_{n m};k)+\nn\\
\fl\qquad
+\frac{1}{4\pi}
\ln\left(\frac{\beta^2}{4 t_{Ew}}\right)+G^{(2)}(\mathbf{R},\mathbf{R};k).
\label{eq::xiewald}
\end{eqnarray}
with $G_f^{(1)}$ to be computed numerically from the integral (\ref{eq:Gf1}), and $G^{(2)}$ from the sum (\ref{eq:G2}). Now we can compute the perturbed part of the spectrum from the condition $\xi_\beta(\textbf{R},k_n)=0$, see (\ref{eq::xisp2}), using the representation (\ref{eq::xiewald}). This final equation in contrast to (\ref{eq::xisp1}) lost the clearness since it depends on the as yet not defined Ewald parameter $t_{Ew}$. To define it we first consider large values of $k$. Then to avoid exponentially large values of the function $g_0(k^2 t_{Ew})$ as well as exponentially large amplitudes of terms with small numbers $n,\,m$ in the expansion of $G^{(2)}$ we put $t_{Ew}=1/k^2$. Obviously this choice is inappropriate for $k\to 0$, since this would mean to compute a huge number of terms in $G^{(1)}$. So, finally the Ewald parameter can be chosen as
\begin{equation}
t_{Ew}=
\left\{\begin{array}{lll}
1/k^2, & \textrm{if} & k>k_0,\\
1/k_0^2, & \textrm{if} & k\leq k_0,
\end{array}\right.
\end{equation}
where $k_0^2=(\pi/d_x)^2+(\pi/d_y)^2$ is the lowest eigenvalue of the unperturbed system. To investigate the spectral statistics we can assume $t_{Ew}=1/k^2$. Then  (\ref{eq::xisp2}) reads
\begin{eqnarray}
\fl
\frac{1}{4\pi}[g_0(1)+\gamma]+\nn\\
\fl\qquad
+\sum_{n,m=-\infty}^\infty \sum_{s_1,s_2=0}^1 (1-\delta_{n,0}\delta_{m,0}\delta_{s_1,0}\delta_{s_2,0})
(-1)^{s_1+s_2}G_f^{(1)}(\mathbf{R},\mathbf{R}_{s_1 s_2}+\mathbf{R}_{n m};k)+\nn\\
\fl\qquad
+\frac{1}{2\pi}
\ln\left(\frac{k\beta}{2}\right)+
G^{(2)}(\mathbf{R},\mathbf{R};k)=0.
\label{eq::xisp3}
\end{eqnarray}
Now (\ref{eq::xisp3}) resembles (\ref{eq::xisp1}), so the main conclusions made above could be repeated. The equation (\ref{eq::xisp3}) is alike (3) in \cite{bog01}, apart from the fact that $G^{(2)}(\mathbf{R},\mathbf{R};k)$ is not a finite sum and the rest in (\ref{eq::xisp3}) is not a polynomial as a function of $k^2$. Equation (\ref{eq::xisp3}) is exact and especially fits for the numerical study, since it contains exponentially convergent series.

For large $k$ the double sum in (\ref{eq::xisp3}) can be neglected and the rest looks very similar to the ``\textit{N}-poles'' approximation \cite{tud08}. Indeed in this regime the spectrum of the billiard can be found from the equation
\begin{eqnarray}
G^{(2)}(\mathbf{R},\mathbf{R};k)+
\frac{1}{2\pi}\left[\ln\left(\frac{k\beta}{2}\right)+\frac{\gamma+g_0(1)}{2}\right]=0,
\label{eq::xisp4}
\end{eqnarray}
where only a finite number of terms in the expansion of $G^{(2)}$ can be taken into account due to the exponential convergency.
Figure \ref{fig::pertsp} shows a graphical interpretation of (\ref{eq::xisp3}). In our calculations we found that approximation (\ref{eq::xisp4}) works perfectly above the first resonance already.

\section{Integrated density of states}

\begin{figure}
\begin{center}
 \includegraphics[width=70mm]{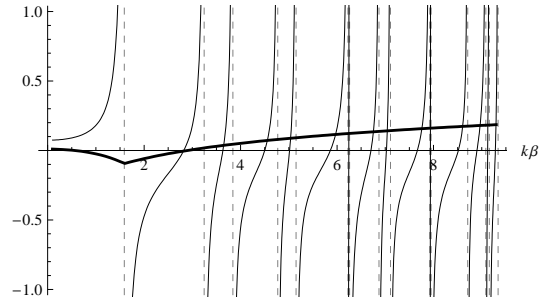}
\end{center}
 \caption{\label{fig::pertsp}Graphical interpretation of (\ref{eq::xisp3}). Vertical grey dashed lines correspond to eigenwavenumbers of the unperturbed billiard. Thin solid line shows $G^{(2)}(\textbf{R},\textbf{R};k)$ as a function of $k$. Thick solid line corresponds to the remainder in (\ref{eq::xisp3}) taken with minus sign.
 In the figure $\beta=1$.}
\end{figure}

In what follows we are interested in level-spacing statistics for the \textit{subset of perturbed eigenvalues} of the \v{S}eba billiard. There are several reasons to restrict ourselves to the statistics of the subspectrum. First of all the influence of the perturbation is more pronounced if one considers only the perturbed part of the spectrum.
This is probably the reason why in the pioneering work \cite{seb90} only the statistics of the subspectrum is considered. Another reason to consider subspectrum's statistics is (\ref{eq::xisp2}), which determines only the perturbed subspectrum. The graphical interpretation of (\ref{eq::xisp3}) gives already an  idea on the structure of the perturbed subspectrum (see figure \ref{fig::pertsp}), while considering the unperturbed subspectrum as well we loose the clearness. The last reason to consider the statistics of the perturbed subspectrum only is the direct correspondence of the perturbed subspectrum to the spectrum obtained from the reflection measurement with a single antenna introduced at the point of the perturbation. In such an experiment the unperturbed subspectrum is not seen at all since the corresponding eigenstates, vanishing at the perturbation point, can not be excited.

\begin{figure}
\begin{center}
 \includegraphics[width=70mm]{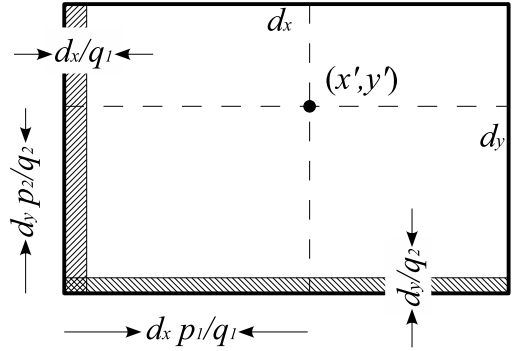}
\end{center}
 \caption{\label{fig::nexp}``Subbilliards'' corresponding to rational ratios $x'/d_x$, $y'/d_y$. The point $(x',y')$ is shown by the black disk.}
\end{figure}

If one considers only the perturbed subspectrum it makes a difference whether the ratios $x'/d_x$ and $y'/d_y$ are rational or irrational numbers (see figure \ref{fig::nexp}). The difference arises from the fact that for irrational numbers all eigenfunctions are perturbed while for rational ones a part of eigenfunctions remains unperturbed.

To compute the statistics and compare it with GOE, semi-Poissonian and Poissonian predictions one should first unfold the spectrum to a mean level spacing of one. This can be achieved by the following definition of the scaled eigenvalues:
\begin{equation}
E_n^{(s)}=N(k_n^2),
\end{equation}
where $N(z)$ is a smoothed function counting a total number of eigenvalues $k_n^2$ less then $z$, i.\,e. the integrated density of states. If a spectrum of a system is known, the function $N(z)$ can be obtained from a numerical fit. For a conventional unperturbed two-dimensional billiard one can use the Weyl estimation of the integrated density of states (see e.\,g. \cite{eck88})
\begin{equation}
N_W(z)=\frac{A_1}{4\pi}z-\frac{A_2}{4\pi}\sqrt{z}+A_W,
\label{eq::nw}
\end{equation}
where $A_1$ is the area of the billiard, $A_2$ is its circumference and $A_W$ is a constant. For the unperturbed rectangular billiard with the sides $d_x,\,d_y$ we obtain $A_1=d_x d_y$ and $A_2=2(d_x+d_y)$.
While the Weyl estimation holds for the whole spectrum of the unperturbed billiard it can not be directly applied to its subspectrum as well as to the perturbed subspectrum of the \v{S}eba billiard. However to fit numerically the integrated density of states one can still assume that the function to be found has the form (\ref{eq::nw}) with some unknown coefficients $A_1,\,A_2,\,A_W$.

The scaled level spacing corresponding to the nearest eigenvalues $k_n^2$ and $k_{n+1}^2$ is
\begin{equation}
s_n=E^{(s)}_{n+1}-E^{(s)}_{n}=N(k_{n+1}^2)-N(k_n^2).
\end{equation}
Thus the constant term $A_W$ in (\ref{eq::nw}) does not influence the statistics. The mean level spacing
\begin{equation}
\langle s\rangle=\frac{1}{M}\sum_{n=1}^M s_n=\frac{1}{M}[N(k_{M+1}^2)-N(k_1^2)]\to 1
\end{equation}
when $M\to\infty$ in accordance with the rescaling requirement.

Though the integrated density of states corresponding to the perturbed subspectrum can be fitted numerically, it is possible to estimate it a priori. Indeed in figure \ref{fig::pertsp} one sees that between two successive eigenvalues of the unperturbed spectrum corresponding to poles of the function $G^{(2)}$ there always exists an eigenvalue of the perturbed billiard. Thus the number of perturbed eigenvalues of the billiard below $z$ should coincide (up to a single eigenvalue) with a number of eigenvalues of the unperturbed billiard below $z$ corresponding to nonvanishing eigenfunctions at the point of the perturbation. Obviously this result does not depend on the value of the scattering length.

Following the argument given above we can compute the expected function $N_e(z)$ right from the unperturbed billiard, where $N_e(z)$ is equal to the integrated density of those states whose eigenfunctions do not vanish at the perturbation point $(x',y')$. Let us assume that $x'=d_x p_1/q_1$, $y'=d_y p_2/q_2$, where $p_1/q_1$ and $p_2/q_2$ are irreducible fractions. From figure~\ref{fig::nexp} one can draw the conclusion that eigenfunctions of the small hatched ``subbilliards'' with Dirichlet conditions at all boundaries are eigenfunctions of the initial billiard and vanish at the point $(x',y')$. According to the Weyl formula the number of eigenvalues below $z$ can be estimated as
\begin{eqnarray}
N_v(z)=\frac{d_xd_y}{4\pi q_1}z-\frac{d_x/q_1+d_y}{2\pi}\sqrt{z}+A_v,\label{eq::nv}\\
N_h(z)=\frac{d_xd_y}{4\pi q_2}z-\frac{d_x+d_y/q_2}{2\pi}\sqrt{z}+A_h\label{eq::nh}
\end{eqnarray}
for the vertical and horizontal hatched billiards respectively. Here $A_v$ and $A_h$ are some constants. We have computed twice the eigenvalues of the billiard obtained as an intersection of these subbilliards. Its number of eigenvalues can be estimated as
\begin{equation}
N_{vh}(z)=\frac{d_xd_y}{4\pi q_1 q_2}z-\frac{d_x/q_1+d_y/q_2}{2\pi}\sqrt{z}+A_{vh}\label{eq::nvh}.
\end{equation}
Finally, the number of eigenvalues corresponding to vanishing eigenfunctions is
\begin{eqnarray}
N_v(z)+N_h(z)-N_{vh}(z)\nn\\
=\left(\frac{1}{q_1}+\frac{1}{q_2}-\frac{1}{q_1q_2}\right)\frac{d_x d_y}{4\pi}z-\frac{d_x+d_y}{2\pi}\sqrt{z}
+A_v+A_h-A_{vh}.
\end{eqnarray}
Subtracting the last estimation from the total number of eigenvalues below $z$
\begin{equation}
N_W(z)=\frac{d_xd_y}{4\pi}z-\frac{d_x+d_y}{2\pi}\sqrt{z}+A_W
\label{eq::nw1}
\end{equation}
we obtain the following estimation for the number of eigenvalues corresponding to nonvanishing eigenfunctions:
\begin{equation}
N_e(z)=\left(1-\frac{1}{q_1}-\frac{1}{q_2}+\frac{1}{q_1q_2}\right)\frac{d_x d_y}{4\pi}z+A_e,
\label{eq::ne}
\end{equation}
where $A_e=A_W+A_{vh}-A_v-A_h$. Surprisingly the surface contribution $\sim\sqrt{z}$ vanishes.

\begin{figure}
\begin{center}
\begin{minipage}{60mm}
\begin{tabular}{c}
 \includegraphics[width=55mm]{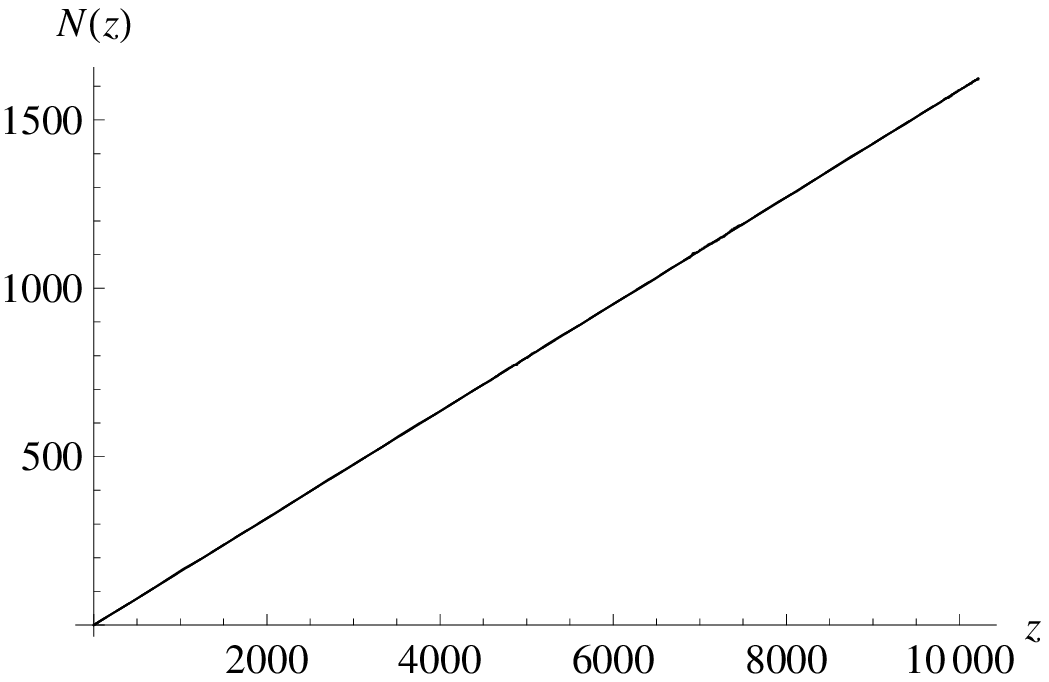} \\
 (a)
\end{tabular}
\end{minipage}
\begin{minipage}{60mm}
\begin{tabular}{c}
 \includegraphics[width=55mm]{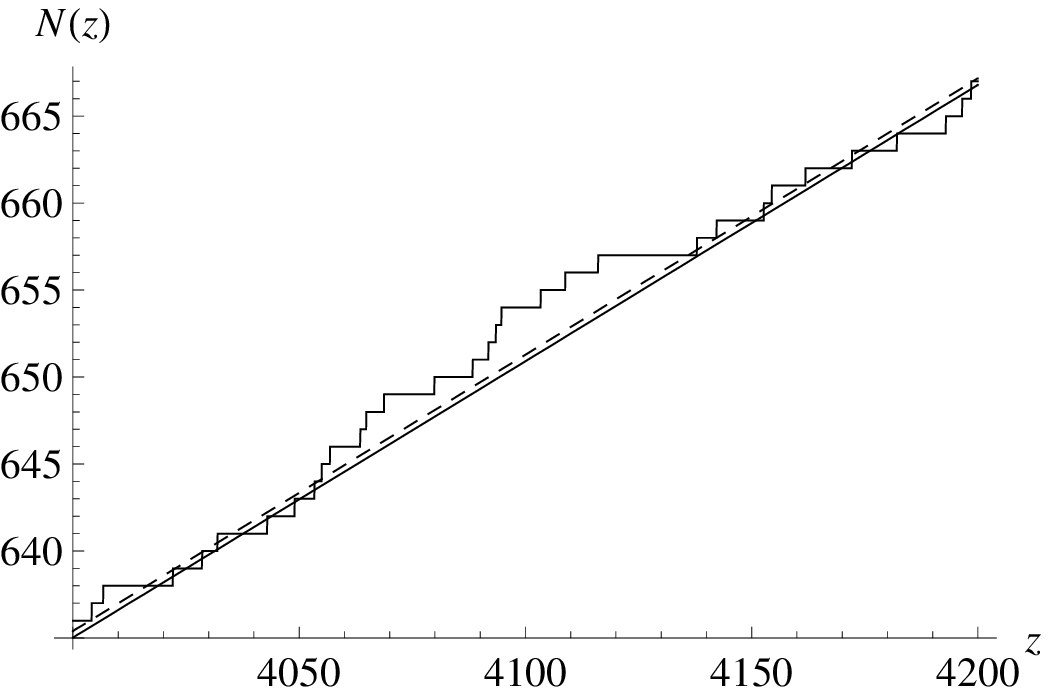} \\
 (b)
\end{tabular}
\end{minipage}
\end{center}
\caption{\label{fig::intdens}Integrated density of states of the perturbed subspectrum of the \v{S}eba billiard. Figure (b) represents the zoom of figure (a). The perturbation with the scattering length $\beta=1$ is placed in the center of the billiard. Stepwise line corresponds to the computed subspectrum, dashed line corresponds to (\ref{eq::ne}), smooth solid line corresponds to the numerical fit of the form (\ref{eq::nw}).}
\end{figure}

In figure \ref{fig::intdens} the estimation (\ref{eq::ne}) and the numerical fit of the form (\ref{eq::nw}) for the integrated density of states are shown for the comparison. One can see that the estimation (\ref{eq::ne}) works very well.

From (\ref{eq::xisp4}), (\ref{eq::ne}) one can compute a shift of a resonance induced by a point perturbation as compared to the mean level spacing. Indeed when $k\to E_{nm}$ the function $G^{(2)}(\mathbf{R},\mathbf{R};k)$ (\ref{eq::g2em}) tends to the following expression:
\begin{equation}
G^{(2)}(\mathbf{R},\mathbf{R};k)\to\frac{4}{d_x d_y}\frac{\sin^2(\pi n x'/d_x)\sin^2(\pi m y'/d_y)}{k^2-E_{nm}}.
\label{eq::g2pole}
\end{equation}
From (\ref{eq::ne}) we find the mean level-spacing $\langle\Delta E\rangle$:
\begin{equation}
\langle\Delta E\rangle=1/N'_e(z)=\frac{4\pi Q}{d_x d_y}, \quad \frac{1}{Q}=1-\frac{1}{q_1}-\frac{1}{q_2}+\frac{1}{q_1q_2}.
\label{eq::mls}
\end{equation}
Substituting (\ref{eq::g2pole}) in (\ref{eq::xisp4}) and using (\ref{eq::mls}) we find for the relative shift of the resonance:
\begin{equation}
\frac{k^2-E_{nm}}{\langle\Delta E\rangle}=\frac{d_x d_y}{4\pi Q}(k^2-E_{nm})=
-\frac{4\sin^2(\pi n x'/d_x)\sin^2(\pi m y'/d_y)}{Q[\ln(E_{nm}\beta^2/4)+\gamma+g_0(1)]}.
\label{eq::resshift}
\end{equation}
Let us estimate the number of resonances needed to show the transition to the Poissonian level-spacing statistics. Then the relative shift should be very small for all sufficiently large numbers $n$ and $m$. The sufficient condition is
\begin{equation}
\frac{Q}{4}[\ln(E_{nm}\beta^2/4)+\gamma+g_0(1)]\gg 1.
\end{equation}
Depending on the values of $Q$ and $\beta$ the value of $E_{nm}$ can be very large.

\section{Level-spacing statistics}

\begin{figure}
\begin{minipage}{80mm}
\begin{tabular}{c}
 \includegraphics[width=80mm]{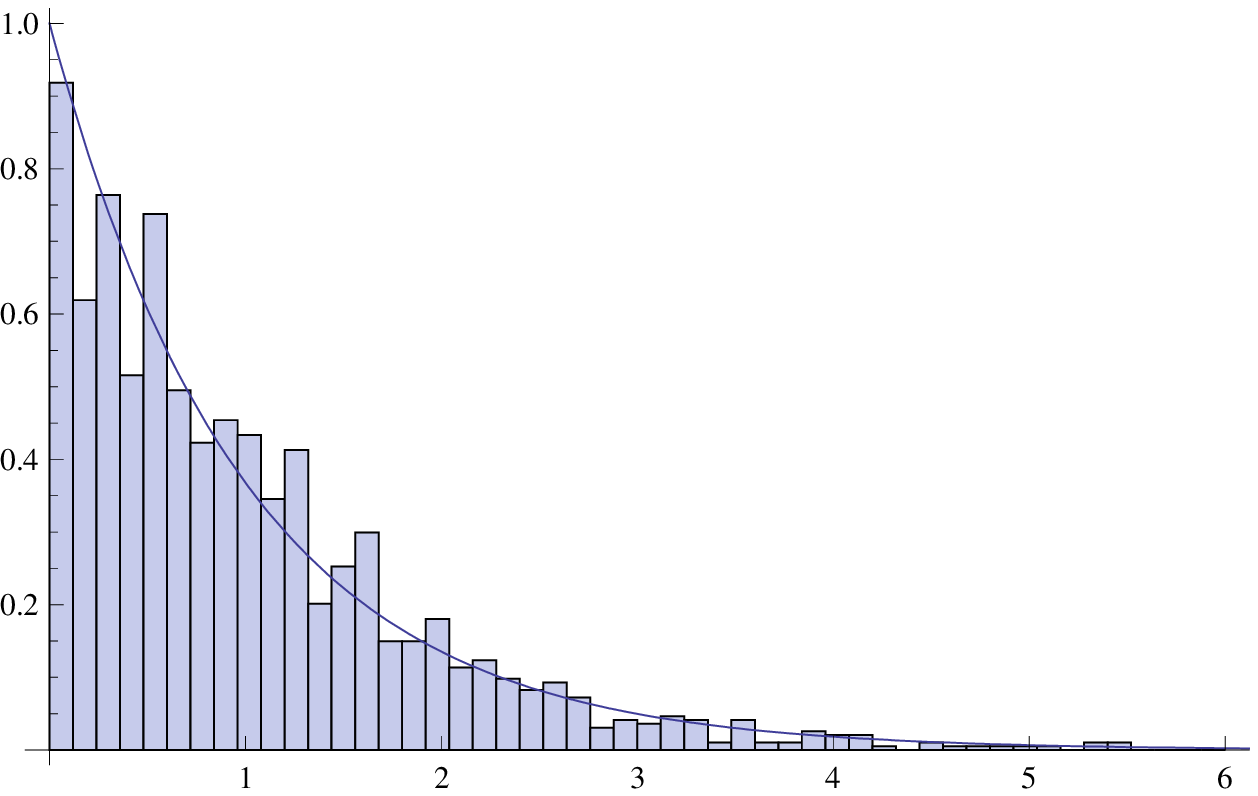} \\ (a)
\end{tabular}
\end{minipage}
\begin{minipage}{80mm}
\begin{tabular}{c}
 \includegraphics[width=80mm]{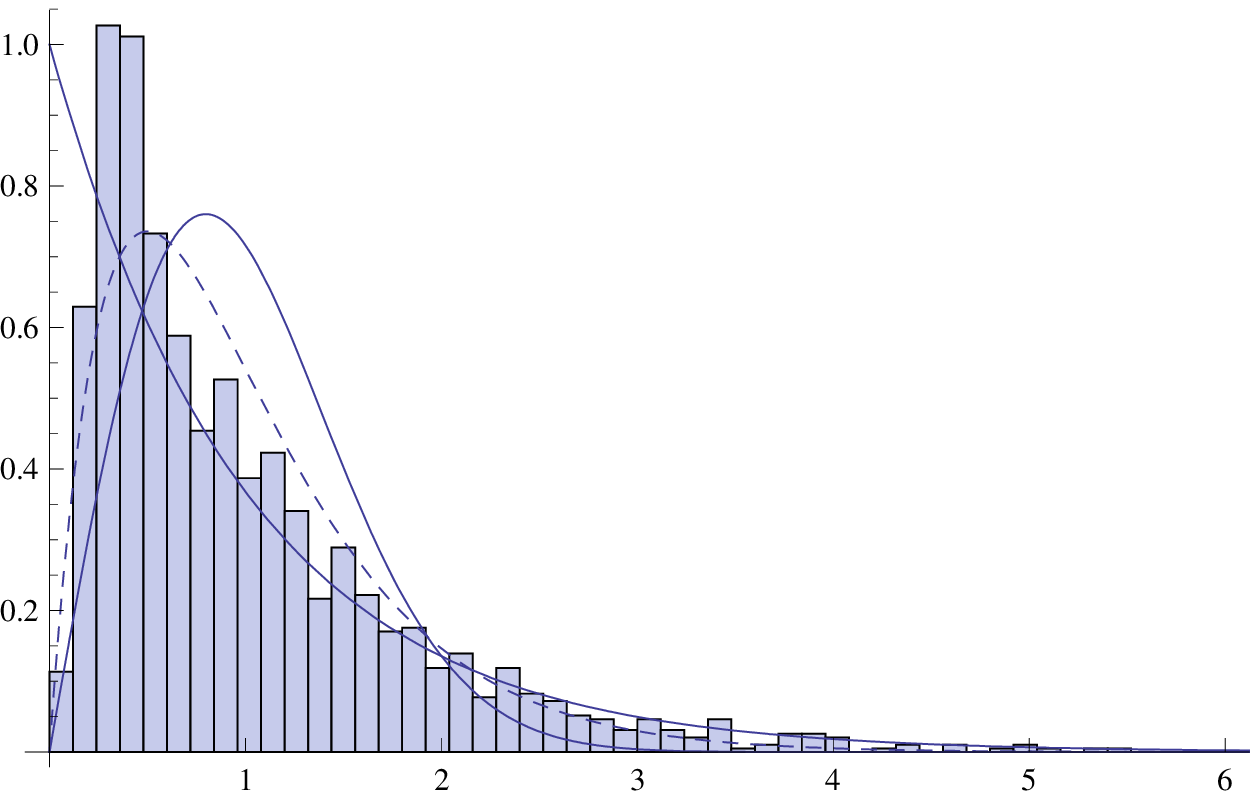} \\ (b)
\end{tabular}
\end{minipage}
\begin{minipage}{80mm}
\begin{tabular}{c}
 \includegraphics[width=80mm]{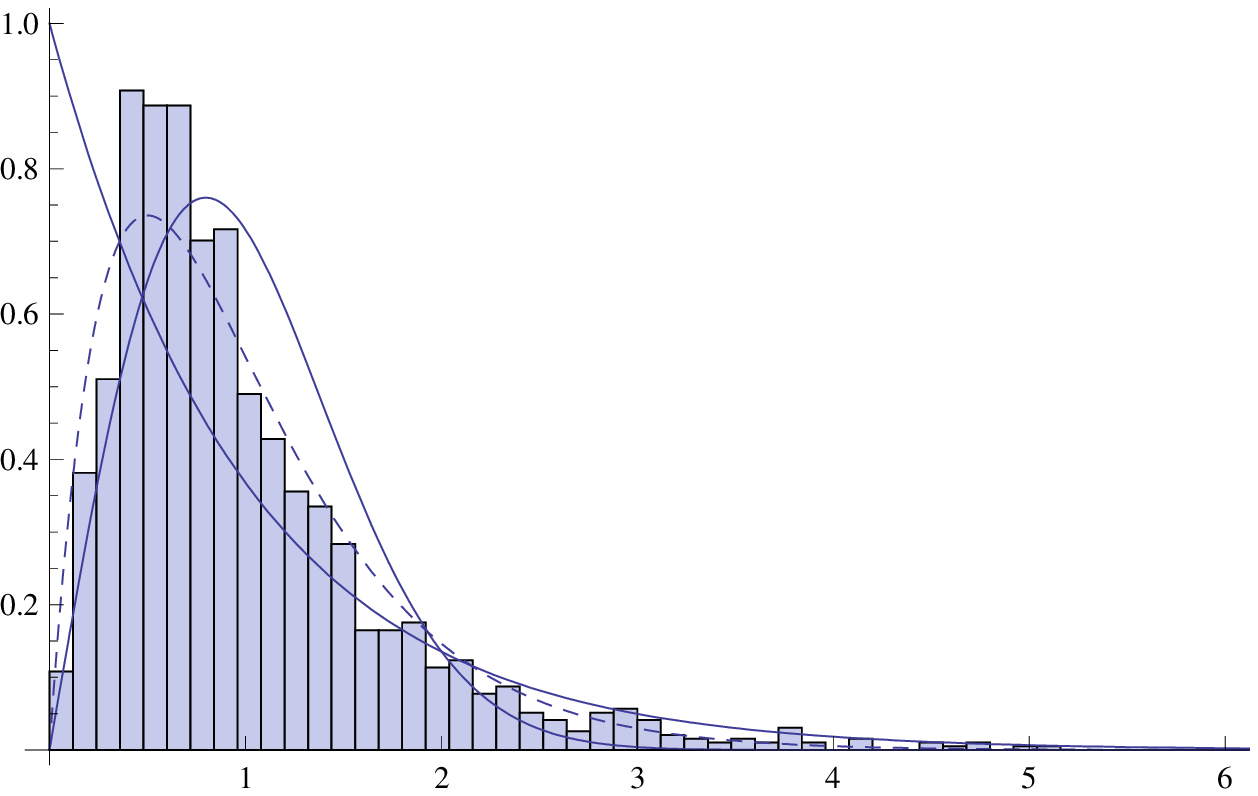} \\ (c)
\end{tabular}
\end{minipage}
\begin{minipage}{80mm}
\begin{tabular}{c}
 \includegraphics[width=80mm]{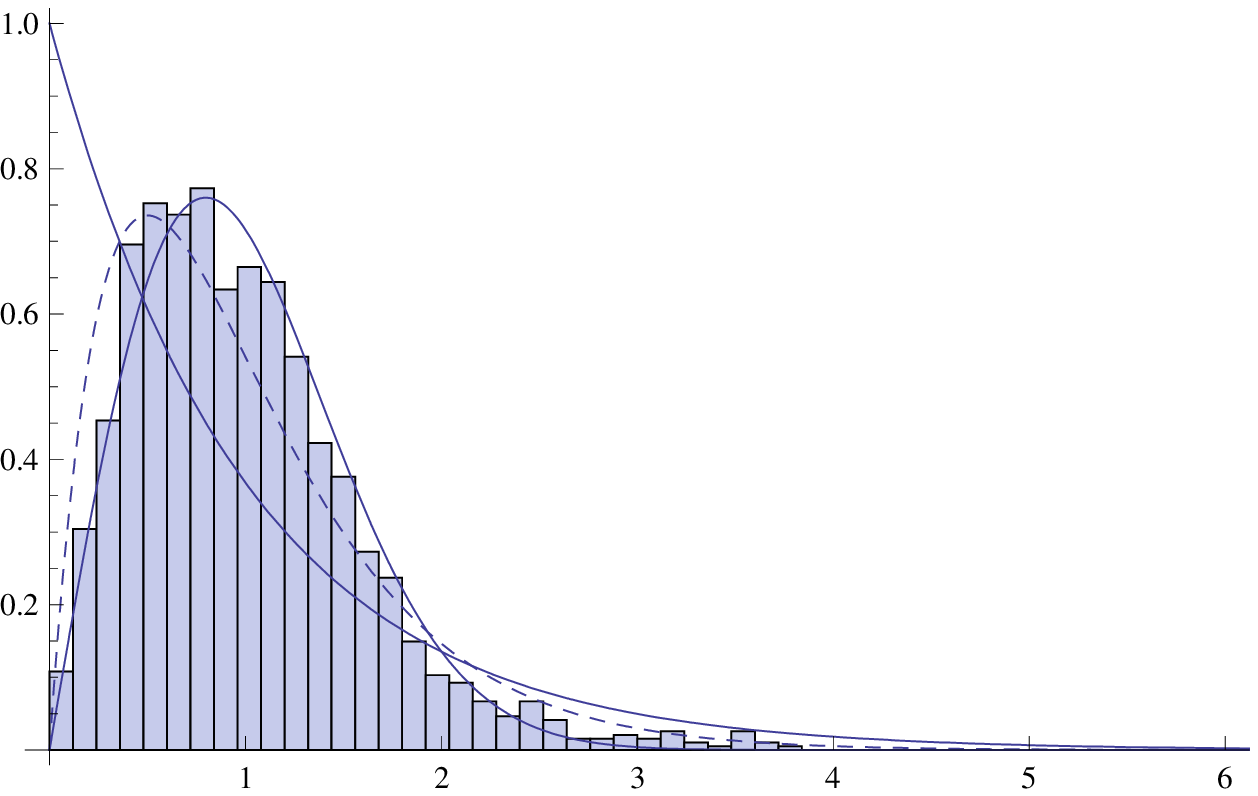} \\ (d)
\end{tabular}
\end{minipage}
 \caption{\label{fig::seba1}The level spacings statistics for the first 1617 resonances. The subspectrum of the unperturbed billiard corresponding to nonvanishing eigenfunctions at the center of the billiard (a). Perturbed subspectrum of the \v{S}eba billiard with a perturbation placed at the center, $\beta=1$ (b). (c) the same as in (b) with $\beta=0.1$. (d) the same as in (b) with $\beta=.02$.}
\end{figure}

\begin{figure}
\begin{minipage}{80mm}
\begin{tabular}{c}
 \includegraphics[width=80mm]{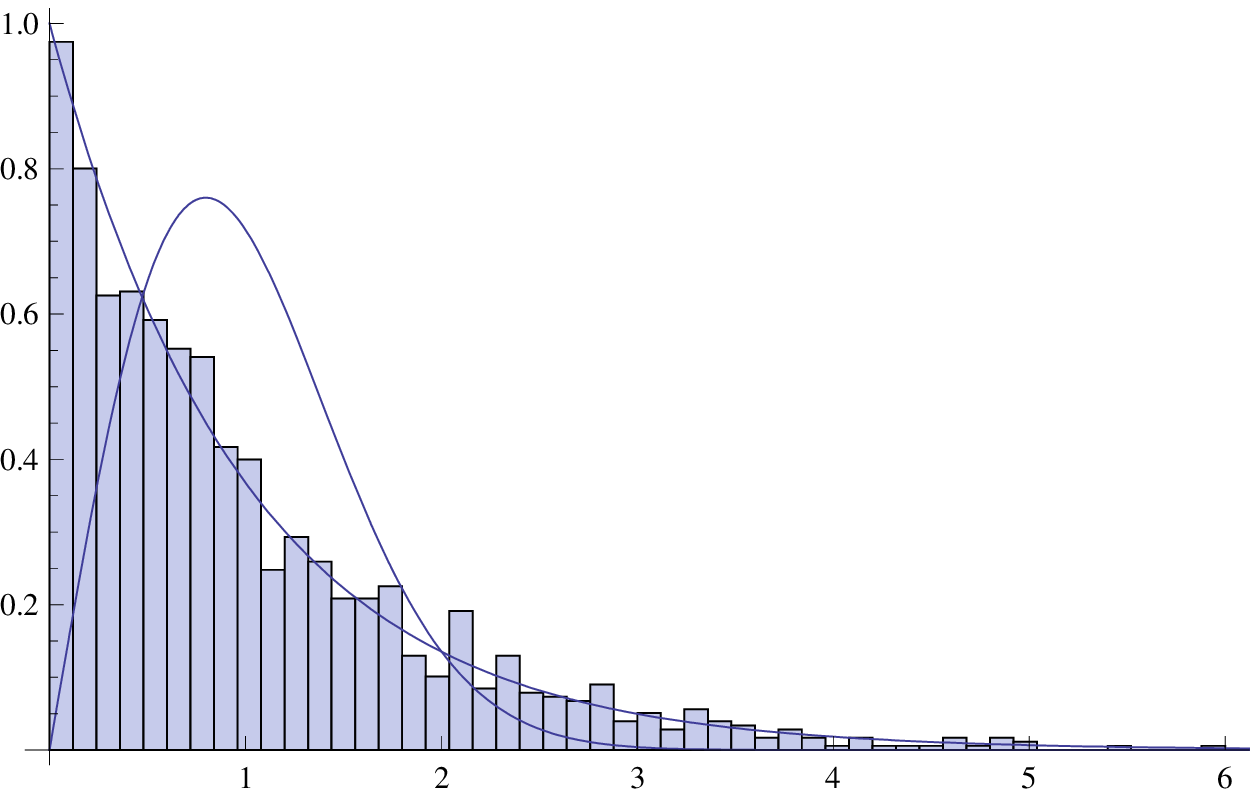} \\ (a)
\end{tabular}
\end{minipage}
\begin{minipage}{80mm}
\begin{tabular}{c}
 \includegraphics[width=80mm]{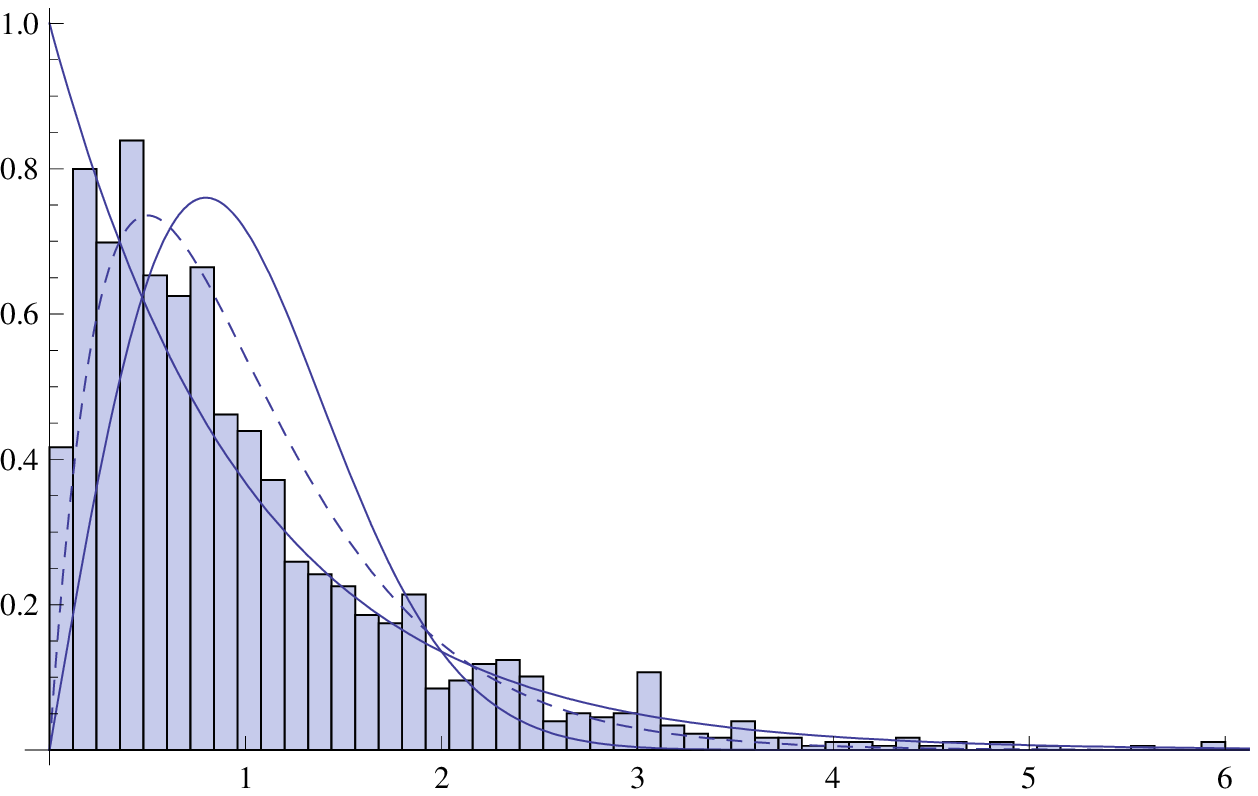} \\ (b)
\end{tabular}
\end{minipage}
\begin{minipage}{80mm}
\begin{tabular}{c}
 \includegraphics[width=80mm]{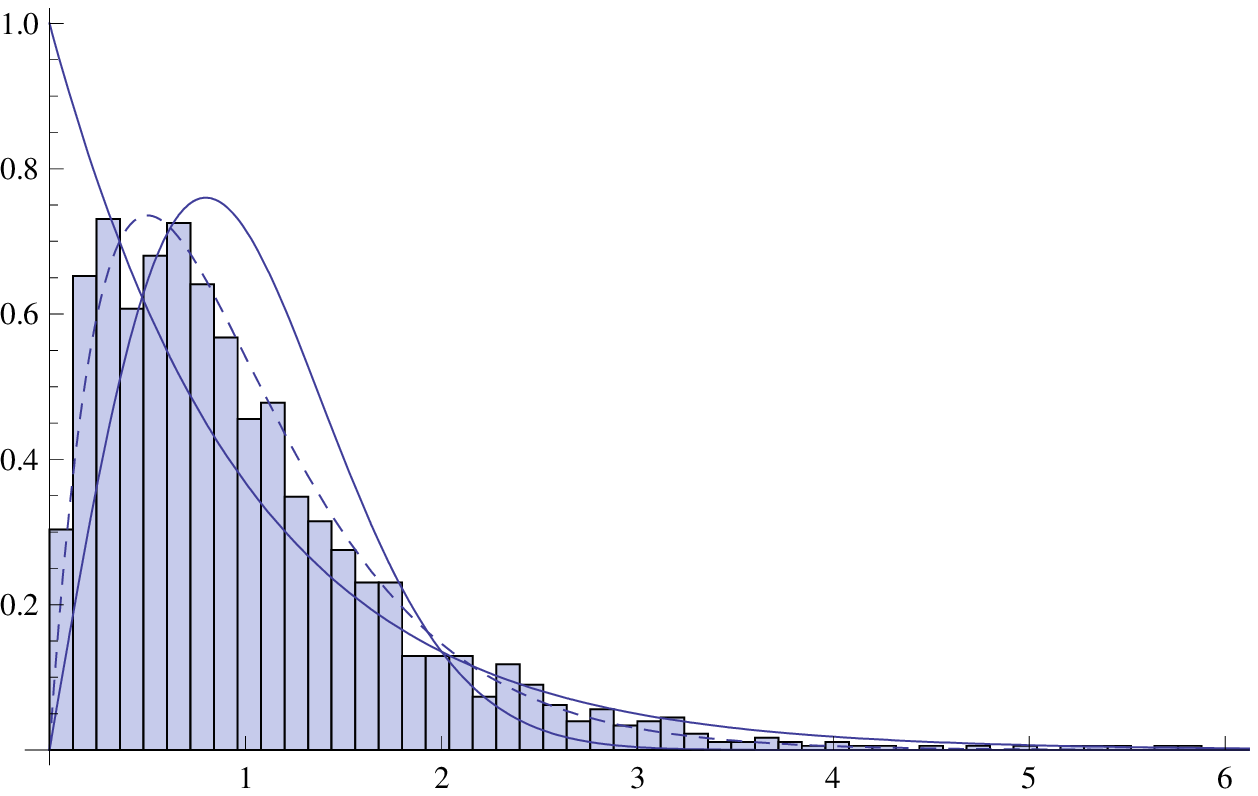} \\ (c)
\end{tabular}
\end{minipage}
\begin{minipage}{80mm}
\begin{tabular}{c}
 \includegraphics[width=80mm]{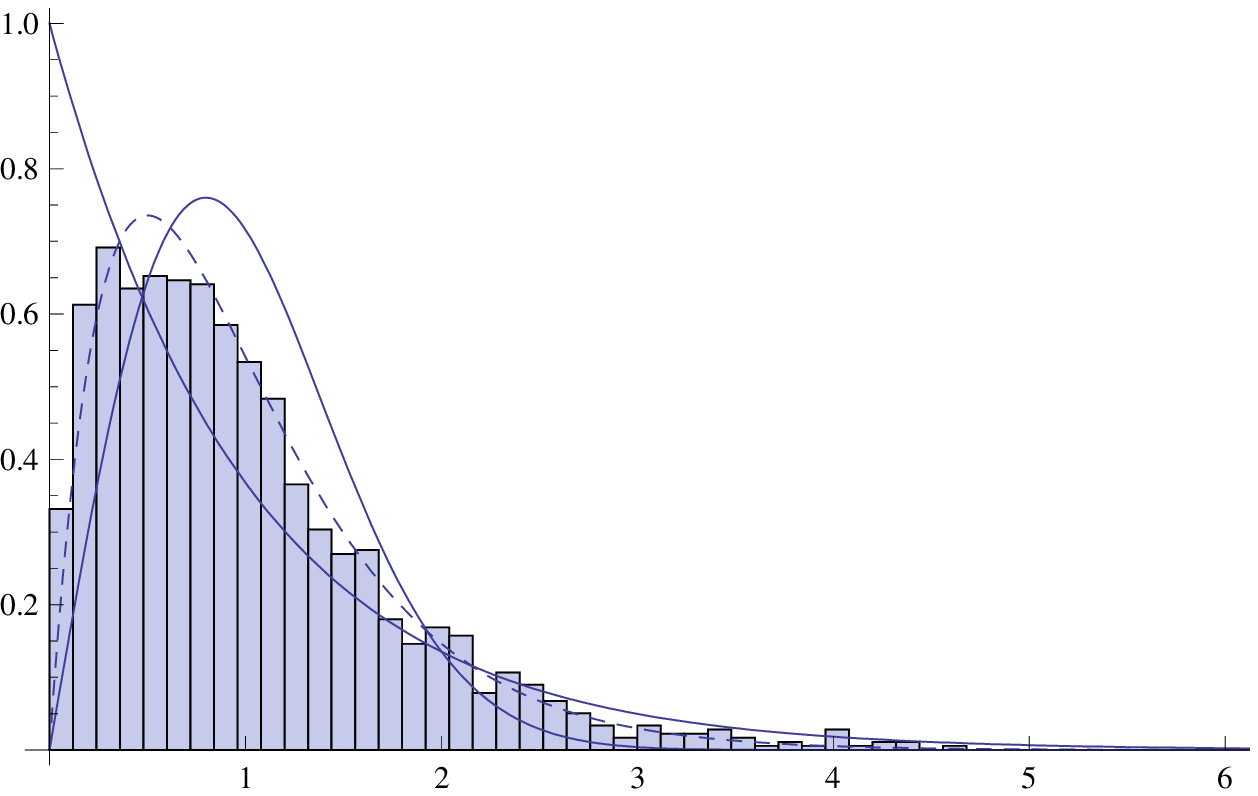} \\ (d)
\end{tabular}
\end{minipage}
 \caption{\label{fig::seba2}The level spacings statistics for the first 1482 resonances. The subspectrum of the unperturbed billiard corresponding to nonvanishing eigenfunctions at the position $(0.55d_x,0.65d_y)$ of the billiard (a). Perturbed subspectrum of the \v{S}eba billiard with a perturbation placed at the point $(0.55d_x,0.65d_y)$, $\beta=1$ (b). (c) the same as in (b) with $\beta=0.1$. (d) the same as in (b) with $\beta=.02$.}
\end{figure}

\begin{figure}
\begin{minipage}{80mm}
\begin{tabular}{c}
 \includegraphics[width=80mm]{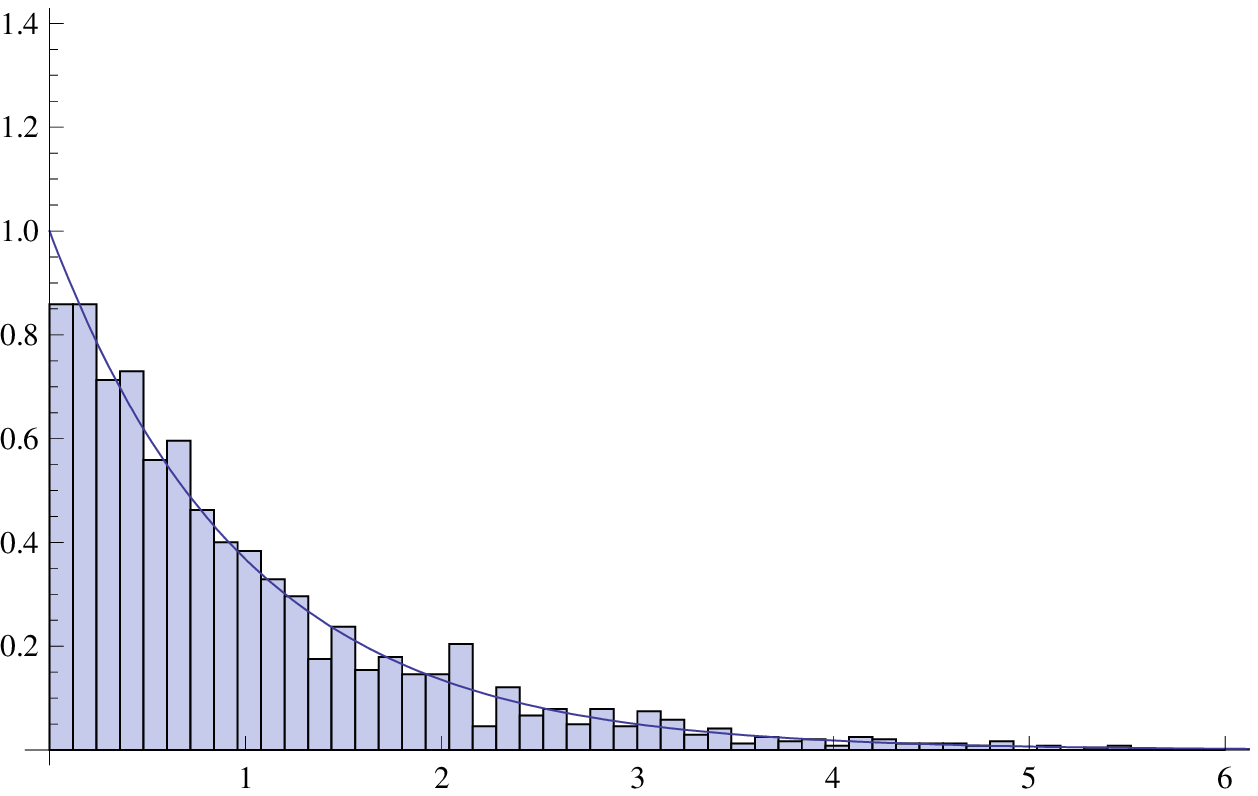} \\ (a)
\end{tabular}
\end{minipage}
\begin{minipage}{80mm}
\begin{tabular}{c}
 \includegraphics[width=80mm]{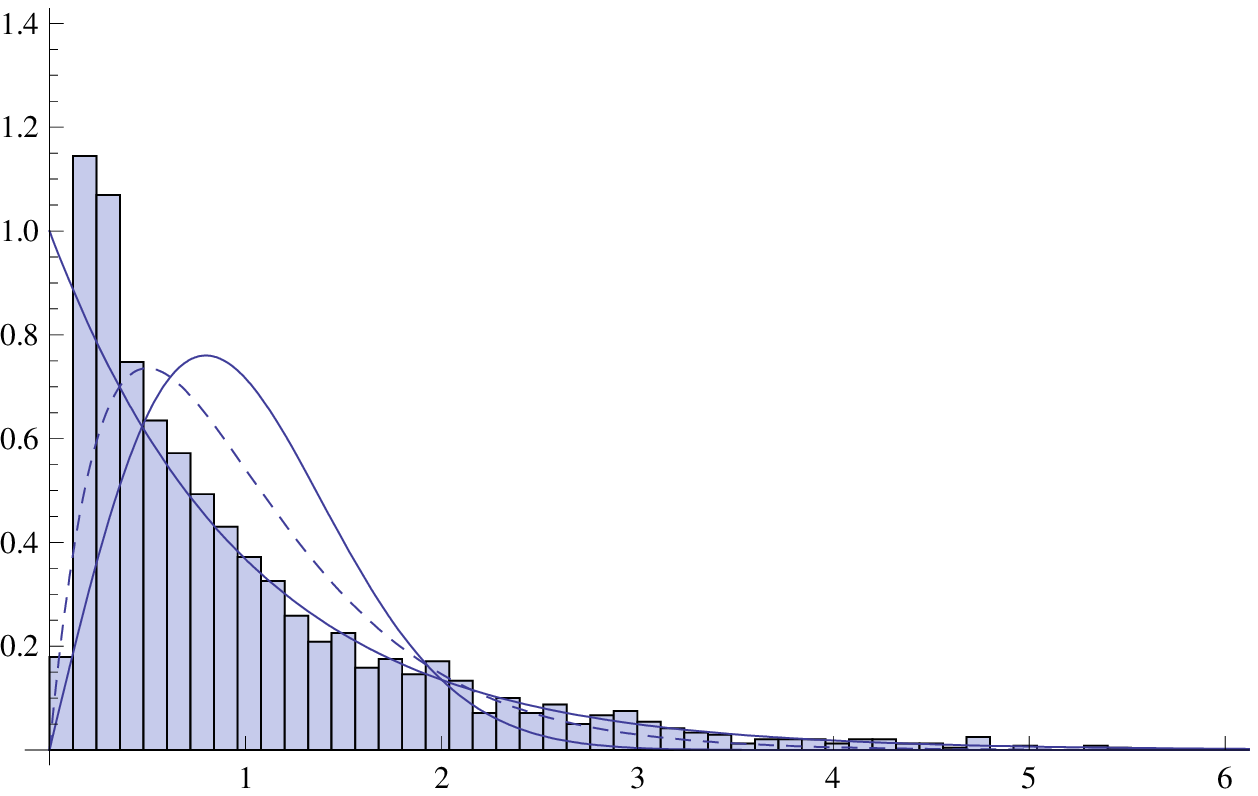} \\ (b)
\end{tabular}
\end{minipage}
\begin{minipage}{80mm}
\begin{tabular}{c}
 \includegraphics[width=80mm]{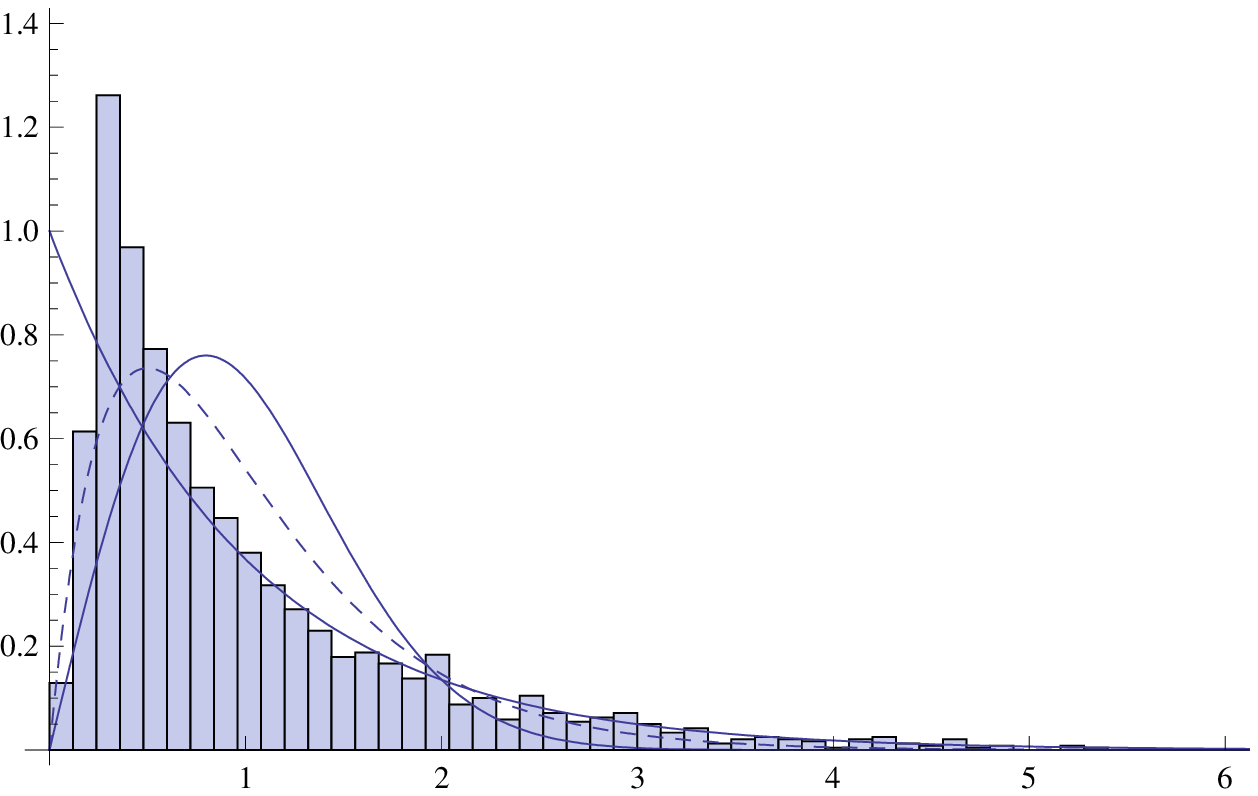} \\ (c)
\end{tabular}
\end{minipage}
\begin{minipage}{80mm}
\begin{tabular}{c}
 \includegraphics[width=80mm]{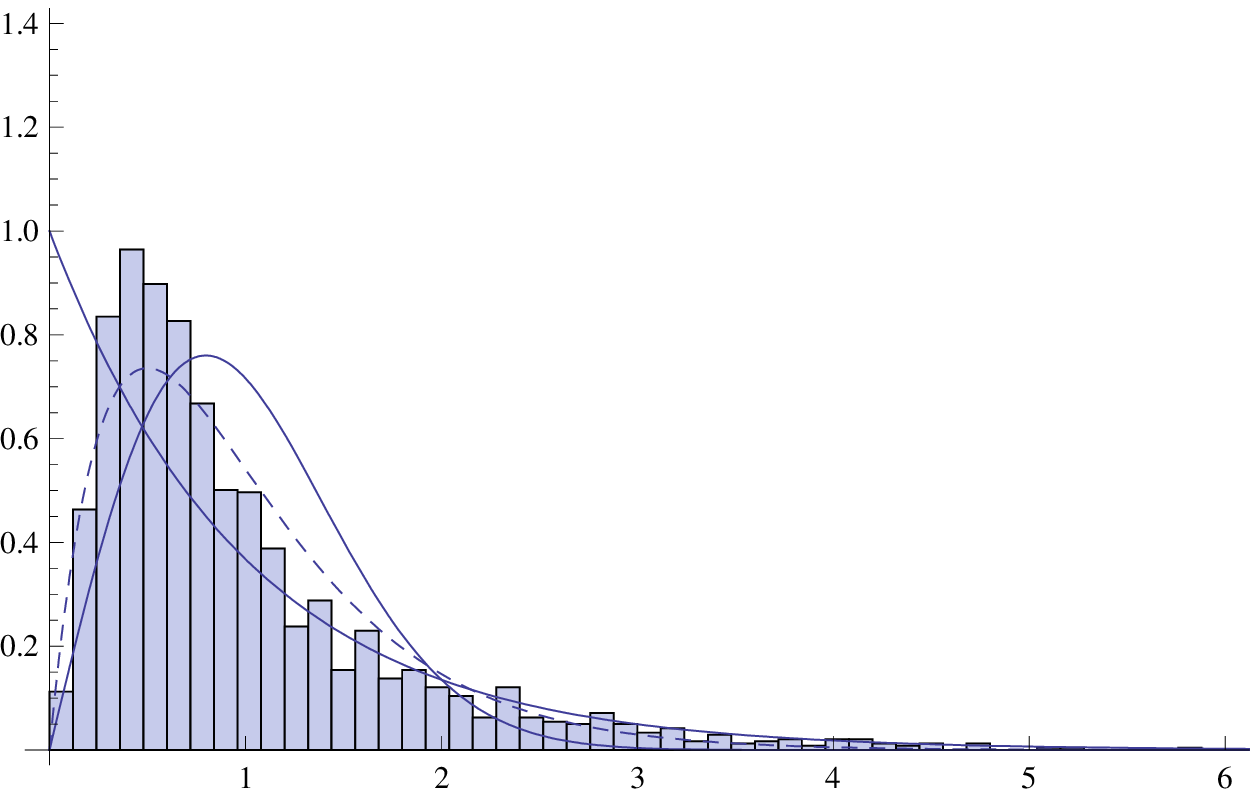} \\ (d)
\end{tabular}
\end{minipage}
 \caption{\label{fig::seba3}The level spacings statistics for the resonances 25000-27000. The subspectrum of the unperturbed billiard corresponding to nonvanishing eigenfunctions at the center of the billiard (a). Perturbed subspectrum of the \v{S}eba billiard with a perturbation placed at the center, $\beta=1$ (b). (c) the same as in (b) with $\beta=0.1$. (d) the same as in (b) with $\beta=.02$.}
\end{figure}

In this section we are presenting a number of numerical results. Figures \ref{fig::seba1}-\ref{fig::seba3} show level-spacings distributions for the perturbed part of the spectrum for various situations to be discussed in detail below. For comparison the curves corresponding to Poissonian, semi-Poissonian and GOE  distributions are plotted by solid, dashed, and solid lines, respectively. In each of the figures subfigure (a) shows the level-spacings distribution for the unperturbed system to make sure that the distribution is really Poissonian, since it is well-known that there may be deviations for small distances depending on the side ratio of the rectangle. Subfigures (b)-(d) show level-spacings distributions for $\beta=1,\,0.1,\,0.02$ respectively. Note that a decreasing value of $\beta$ means an increase of the perturbation.

In figure \ref{fig::seba1} the scatterer is placed in the center whereas in figure \ref{fig::seba2} it is at the point $(0.55d_x,0.65d_y)$. In both cases about 1500 lowest perturbed eigenvalues have been considered. For $\beta=1$, i.e. for a weak perturbation, the distribution shows a linear repulsion for small distances and an exponential tail, but not a semi-Poissonian behavior in the strict sense (dashed line). With $\beta=0.1,\,0.02$ there is a gradual transition to a broader distribution, resembling GOE one for the scatterer in the center (figure \ref{fig::seba1}(d)\,) and a semi-Poissonian distribution for the scatterer in the off-center position (figure \ref{fig::seba2}(d)\,). This observation would deserve more quantitative treatment, but this goes beyond the scope of this paper. Qualitatively it may be understood from the fact that in the center of the billiard all perturbed eigenfunctions have the same value, while for an off-center position there is a distribution of eigenfunctions amplitudes giving rise to a corresponding distribution of resonances shifts.

Figure \ref{fig::seba3} finally shows the level-spacings distribution again with the perturbation in the center but now for the numbers of perturbed eigenvalues from 25000 to 27000. Comparison of figures \ref{fig::seba1} and \ref{fig::seba3} shows a pronounced change of the distribution towards Poissonian with increasing eigenvalues numbers. This is particularly evident for the weaker perturbations $\beta=1,\,0.1$ (figure \ref{fig::seba3}(b)-(c) and demonstrates the main result of this paper: \textit{with increasing eigenvalues numbers eventually the level-spacings distribution of the unperturbed system is recovered.}

\section{Conclusions}

Let us now conclude. First of all we have presented in the paper the complete solution of the spectral problem for the rectangular billiard with a single point perturbation. In contrast to previous studies \cite{seb90, bog01} we have shown that the statistics of the \v{S}eba billiard tends to a Poissonian when the number of levels taken into account tends to infinity. The estimation given at the end of Section 5 showed, however, that the transition to Poissonian statistics appears, depending on the scattering length,  only at exponentially large quantum numbers. The solution is based on the \textit{Ewald representation of the renormalized Green function} (\ref{eq::xiewald}). This representation contains exponentially rapidly convergent series. Together with the Ewald representation of the usual Green function (\ref{eq::g1f-uv}), (\ref{eq::g1}), (\ref{eq::g2em}) the presented approach is a powerful tool to analyze various experiments made in rectangular billiards.

\ack
The authors gratefully acknowledge useful discussions with P.~\v{S}eba, P.~Exner, V.~G.~Papanicolaou, N.~Makarov, F.~Izrailev, M. Miski-Oglu, B. Dietz, P.~Kurasov, S.~Fishman and A.~Potzuweit.

This work was supported by the Deutsche Forschungsgemeinschaft via an individual grant.

\appendix
\setcounter{section}{1}
\section*{Appendix}

In the Appendix we show that the explicit application of the Poisson resummation in the derivation of Ewald's representation of the Green function can be avoided. These findings simplify the technical calculations.

Let us consider the initial problem (\ref{eq::ginit}) in the rectangular billiard with proper boundary conditions. Then the solution can be written in two equivalent forms: in the form of the images representation $g_i(t;\mathbf{r},\mathbf{R},k)$ and in the form of eigenmodes representation $g_e(t;\mathbf{r},\mathbf{R},k)$:
\begin{eqnarray}
g_i(t;\mathbf{r},\mathbf{R},k)=
\sum_{n,m=-\infty}^\infty \sum_{s_1,s_2=0}^1 (-1)^{s_1+s_2}g(t;\mathbf{r},\mathbf{R}_{s_1s_2}+\mathbf{R}_{nm},k)\\
g_e(t;\mathbf{r},\mathbf{R},k)=
\frac{4}{d_x d_y}\sum_{n,m=1}^\infty e^{[k^2-(\pi n/d_x)^2-(\pi m/d_y)^2]t}
\psi_{nm}(x,y)\psi_{nm}(x',y').
\end{eqnarray}
Then we write
\begin{eqnarray}
G(\mathbf{r},\mathbf{R};k)=G^{(1)}(\mathbf{r},\mathbf{R};k)+G^{(2)}(\mathbf{r},\mathbf{R};k),
\end{eqnarray}
where
\begin{eqnarray}
G^{(1)}(\mathbf{r},\mathbf{R};k)=-\int_0^{t_{Ew}}g_i(t;\mathbf{r},\mathbf{R},k)dt,\label{eq::g1g}\\
G^{(2)}(\mathbf{r},\mathbf{R};k)=-\int_{C_5}g_e(t;\mathbf{r},\mathbf{R},k)dt\label{eq::g2g}.
\end{eqnarray}
Performing the integration in Eqs.~(\ref{eq::g1g}), (\ref{eq::g2g}) we immediately get (\ref{eq::g1}), (\ref{eq::g2em}).

\vspace{.5cm}

\bibliographystyle{unsrt}
%\bibliography{thispaper}

\begin{thebibliography}{10}

\bibitem{seb90}
P.~\v{S}eba.
\newblock Wave chaos in singular quantum billiard.
\newblock {\em Phys. Rev. Lett.}, 64:1855, 1990.

\bibitem{cas85}
G.~Casati, B.~V. Chirikov, and I.~Guarneri.
\newblock Energy-level statistics of integrable quantum systems.
\newblock {\em Phys. Rev. Lett.}, 54:1350, 1985.

\bibitem{sin70}
Ya.~G. Sinai.
\newblock Dynamical systems with elastic reflections.
\newblock {\em Russian Math. Surveys}, 25:137, 1970.

\bibitem{ber81}
M.~V. Berry.
\newblock Quantizing a classically ergodic system: Sinai's billiard and the
  {KKR} method.
\newblock {\em Ann. Phys. (N.Y.)}, 131:163, 1981.

\bibitem{seb91}
P.~\v{S}eba and K.~{\.{Z}}yczkowski.
\newblock Wave chaos in quantized classically nonchaotic systems.
\newblock {\em Phys. Rev. A}, 44:3457, 1991.

\bibitem{haa91}
F.~Haake, G.~Lenz, P.~\v{S}eba, J.~Stein, H.-J. St\"ockmann, and
  K.~{\.{Z}}yczkowski.
\newblock Manifistation of wave chaos in pseudointegrable microwave resonators.
\newblock {\em Phys. Rev. A}, 44:R6161, 1991.

\bibitem{bog01}
Eug\`{e}ne Bogomolny, Ulrich Gerland, and Charles Schmit.
\newblock Singular statistics.
\newblock {\em Phys. Rev. E}, 63:036206--1, 2001.

\bibitem{dem88}
Yu.~N. Demkov and V.~N. Ostrovskiy.
\newblock {\em Zero-range potentials and their applications in atomic physics}.
\newblock Plenum Press, New York, 1988.

\bibitem{alb88}
S.~Albeverio, F.~Gesztesy, R.~H{\o}egh-Krohn, and H.~Holden.
\newblock {\em Solvable Models in Quantum Mechanics}.
\newblock Springer, New York, Berlin, 1988.

\bibitem{tud08}
T.~Tudorovskiy, R.~H\"ohmann, U.~Kuhl, and H.-J. St\"ockmann.
\newblock On the theory of cavities with point-like perturbations. {P}art {I}:
  General theory.
\newblock {\em J. Phys. A: Math. Theor.}, 41:275101, 2008.

\bibitem{gra80}
I.~S. Gradshtein and I.~M. Rizhik.
\newblock {\em Tables of Integrals, Series and Products}.
\newblock Academic Press, Orlando, 1980.

\bibitem{exn97}
P.~Exner and P.~$\check{\rm S}$eba.
\newblock Resonance statistics in a microwave cavity with a thin antenna.
\newblock {\em Phys. Lett. A}, 228:146, 1997.

\bibitem{ewa21}
P.~P. Ewald.
\newblock Die {B}erechnung optischer und elektrostatischer {G}itterpotentiale.
\newblock {\em Annalen der Physik}, 64:253, 1921.
\newblock Translated by {A}.{C}ornell, {A}tomics {I}nternational {L}ibrary,
  1964.

\bibitem{lin98}
C.~M. Linton.
\newblock The {G}reen's function for the two-dimensional {H}elmholtz equation
  in periodic domains.
\newblock {\em Journ. Eng. Math.}, 33:377, 1998.

\bibitem{lin99}
C.~M. Linton.
\newblock Rapidly convergent representations for {G}reen's functions for
  {L}aplace's equation.
\newblock {\em {P}roceedings of the {R}oyal {S}ociety A}, page 1767, 1999.

\bibitem{duf01}
Dean~G. Duffy.
\newblock {\em Green's functions with applications}.
\newblock CRC Press, Boca Raton, London, New York, Washington, D.C., 2001.

\bibitem{pap99}
V.~G. Papanicolaou.
\newblock {E}wald's method {R}evisited: {R}apidly convergent series
  representations of certain {G}reen's functions.
\newblock {\em Journ. Comp. Anal. Applications}, 01:105, 1999.

\bibitem{mor06}
Alexander Moroz.
\newblock Quasi-periodic {G}reen's functions of the {H}elmholtz and {L}aplace
  equations.
\newblock {\em J. Phys. A: Math. Gen.}, 39:11247, 2006.

\bibitem{eck88}
B.~Eckhardt.
\newblock Quantum mechanics of classically non-integrable systems.
\newblock {\em Phys. Rep.}, 163:205, 1988.

\end{thebibliography}

\end{document}